\documentclass[superscriptaddress,aps,prb,reprint]{revtex4-1}

\newcommand{\bra}[1]{\langle #1 |}			
\newcommand{\ket}[1]{| #1 \rangle}

\renewcommand{\Re}{\operatorname{Re}}
\renewcommand{\Im}{\operatorname{Im}}

\usepackage{commath}

\usepackage[table]{xcolor}
\usepackage{amsmath,amssymb}
\usepackage[utf8]{inputenc}
\usepackage{graphicx}
\usepackage[caption=false,subrefformat=parens,labelformat=parens]{subfig}
\usepackage{epstopdf}
\usepackage{siunitx}
\usepackage{tikz}
\usepackage{placeins}
\usepackage{cleveref}
\usepackage{bm}

\begin{document}
\begin{abstract}
Group theory and density functional theory methods are combined to obtain compact and accurate $k\cdot p$ Hamiltonians that describe the bandstructures around the $K$ and $\Gamma$ points for the 2D material hexagonal boron arsenide (h-BAs) predicted to be an important low-bandgap material for electric, thermoelectric, and piezoelectric properties that supplements the well-studied 2D material hexagonal boron nitride. Hexagonal boron arsenide is a direct bandgap material with band extrema at the $K$ point. The bandgap becomes indirect with a conduction-band minimum at the $\Gamma$ point subject to a strong electric field or biaxial strain. At even higher electric field strengths (approximately $\SI{0.75}{\volt\per\angstrom}$) or a large strain ($14$~\%) 2D hexagonal boron arsenide becomes metallic. Our $k\cdot p$ models include to leading orders the influence of strain, electric, and magnetic fields. Excellent qualitative and quantitative agreement between density functional theory and  $k\cdot p$ predictions are demonstrated for different types of strain and electric fields. 
\end{abstract}

\title{Electronic structure and tunability of 2D hexagonal boron arsenide}

\author{Mathias Rosdahl Brems}
\affiliation{Technical University of Denmark, Department of Photonics Engineering, Kongens Lyngby, 2800, Denmark}

\author{Morten Willatzen}
\affiliation{Technical University of Denmark, Department of Photonics Engineering, Kongens Lyngby, 2800, Denmark}
\affiliation{Beijing Institute of Nanoenergy and Nanosystems, Chinese Academy of Sciences, Beijing, P. R. China}
\affiliation{School of Nanoscience and Technology, University of Chinese Academy of Sciences, Beijing 100049, P. R. China}

\maketitle

\section{Introduction}

Two-dimensional (2D) materials have opened a new field of research initiated by the micromechanical cleavage of graphene and the characterization of the unique properties of 2D materials\cite{graphene,graphene2}. Since the discovery and isolation of graphene in 2004, a vast number of 2D materials has been synthesized using various fabrication techniques \cite{fabrication_review_ald,fabrication_review_epitaxy,fabrication_review_cvd}. With the increasing number of materials and possibilities of functionalization and combining them in heterostructures\cite{heterostructures_review}, they constitute a versatile class of materials with a large range of possible applications within nanotechnology. Strain engineering is a well established tool to optimize the electronic properties of semiconductors\cite{strainengineering_cmos}, and is particularly relevant for 2D materials due to their ability to sustain much larger strain magnitudes than bulk crystals\cite{strainengineering_dai,strainengineering_akinwande}. Strain magnitudes above $10\%$ can be achieved in graphene without damaging the material\cite{graphene_garza}  and the effect of strain on the bandstructure has been studied theoretically\cite{graphene_winkler,graphene_cocco} and experimentally\cite{graphene_yan,graphene_li}. This gives potential for use in strain-sensing applications\cite{graphene_lee,graphene_wang,graphene_yang}.

\begin{figure}
\centering
\subfloat[ ][]{\includegraphics[width=\columnwidth]{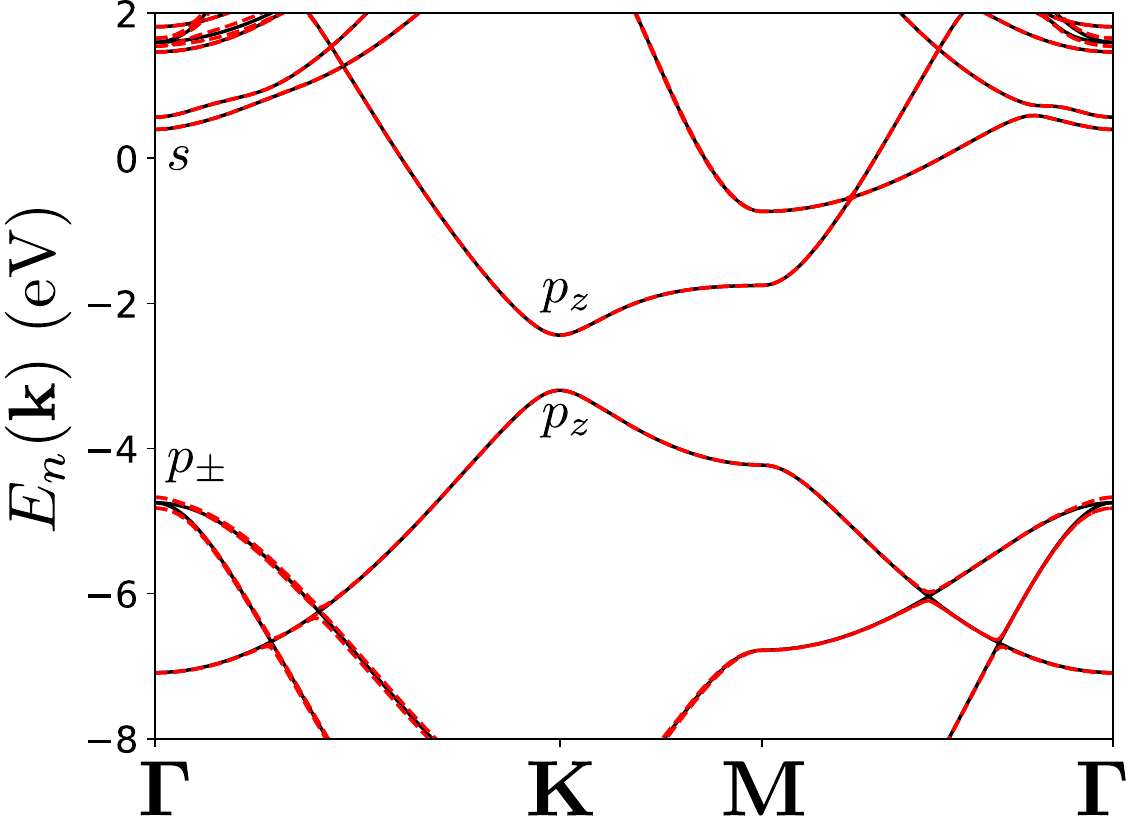}\label{bandstructure}}
\\
\subfloat[ ][]{\includegraphics[width=0.47\columnwidth]{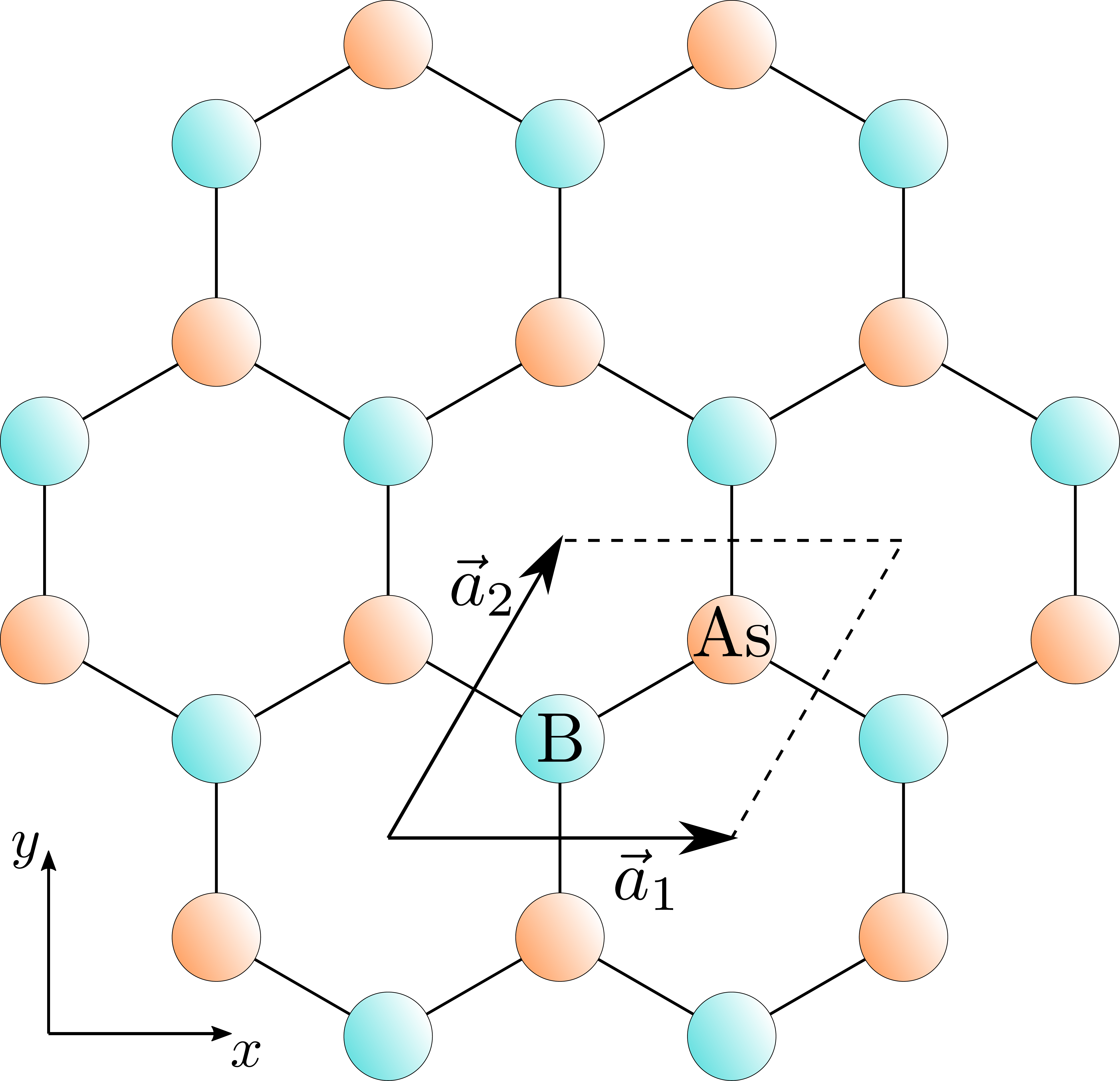}}
\quad
\subfloat[ ][]{\includegraphics[width=0.47\columnwidth]{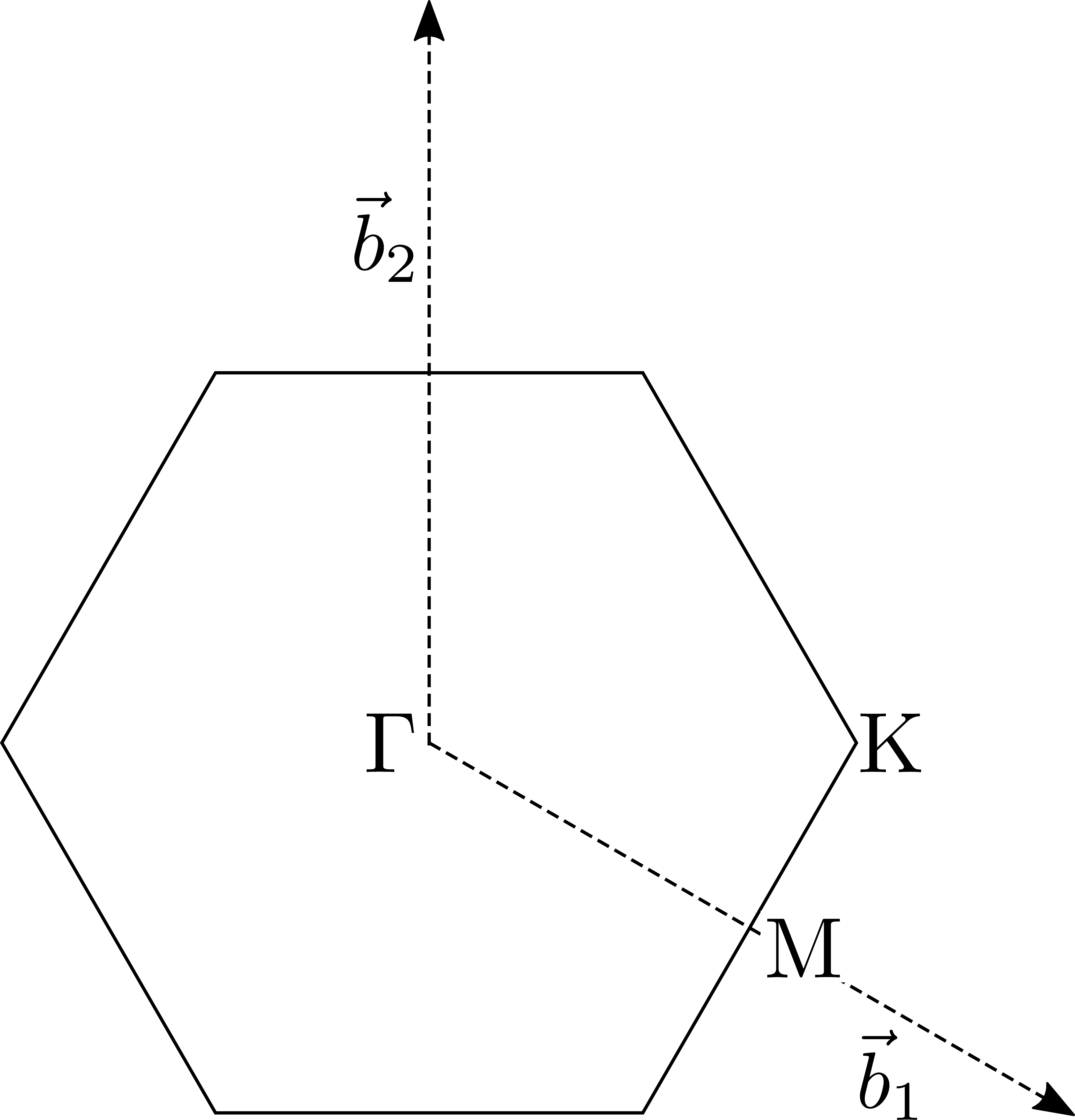}}
\caption{ (a) Electronic band structure of 2D hexagonal boron arsenide (h-BAs) with (red dashed) and without (black solid) SOC and the orbital character of the relevant bands (b) Crystal structure of 2D hexagonal boron arsenide. The bond length after relaxation is $\SI{3.39}{\angstrom}$. (c) First Brillouin zone and reciprocal lattice vectors.}
\label{structure}
\end{figure}

The $k\cdot p$ method has been used to derive effective models to describe the electronic structure of graphene, silicene, phospherene, MoS$_2$ etc. \cite{WillatzenNJP, MoS2_xiao,MoS2_bereinvand,MoS2_kormanyos}. Strain is known to strongly affect the electronic properties of  two-dimensional transition metal dichalcogenides (TMDC)~\cite{MoS2_he,WSe2_schmidt,MoS2_conley,WSe2_wu,MoS2_gomez}, and a recent study shows that for photodetector applications strain can tune the photoresponsivity of MoS$_2$ by 2-3 orders of magnitude\cite{MoS2_gant}. Another way to engineer the electronic properties of 2D materials is by application of a perpendicular electric field\cite{MoS2_zubair,MoS2_qi}.

Computational studies revealed that boron arsenide (BAs) can form a stable 2D hexagonal structure similar to hexagonal boron nitride and explored the material properties using density functional theory, showing a semiconducting nature with a bandgap of approximately $\SI{1}{\electronvolt}$\cite{BAs,BAs_sahin}. Boron arsenide forms a cubic 3D structure with a remarkably high thermal conductivity\cite{BAs_tian,BAs_kang}, a property the 2D structure h-BAs is expected to share\cite{BAs_shi} with application perspectives as coolant in nanodevices. In Ref. \onlinecite{BAs} they explore the effect of biaxial strain on the electronic structure and find a transition to a metallic state at a biaxial strain of 14\%. The application of 2D h-BAs for gas sensing has also been examined\cite{BAs_ren}. Functionalization of h-BAs has been studied theoretically\cite{BAs_zhang,BAs_ullah} and is found promising for applications in e.g. spintronics. Hexagonal BAs is an interesting 2D candidate since it has the same symmetry (point group $D_{3h}$) as the important and well-studied 2D material h-BN. In contrast to h-BN, which is a high-bandgap semiconductor ($5.9$~eV), the bandgap of h-BAs is much smaller yet widely tunable through application of strain and external fields. 

In this work we develop analytical and computationally fast models for the bandstructure of 2D h-BAs based on the $k\cdot p$ method and group theory. We aim to gain a physical understanding of this material as well as a simple accurate model describing the band structure close to the high symmetry points both qualitatively and quantitatively including the spin-orbit coupling (SOC). Standard density functional theory calculations are used to determine the relevant model parameters. We carry out a detailed investigation of strain effects and symmetry-allowed terms extending the study beyond biaxial strain. A perpendicular electric field is found to have a strong influence on the lowest conduction band, giving rise to a semiconductor to metal transition for strong electric fields.

\section{Effective models}
In this section we will derive effective models describing the band structure close to the $\mathbf K$ and $\mathbf \Gamma$ point, including effects of external electric or magnetic fields and strain. The derivation is based on the $k\cdot p$ method\cite{kpmethod,BirPikus} and relies on the symmetry of the crystal as well as information from DFT calculations about the atomic orbitals contributing to the relevant bands. In Fig.~\ref{structure} we show the band structure from DFT, the crystal structure and first Brillouin zone of 2D hexagonal boron arsenide. The point group  is $D_{3h}$, the character table is shown in Table~\ref{chartables}. In Table~\ref{charprodtables} product tables of irreducible representations are given. 

\subsection{Single-particle Hamiltonian in the $k\cdot p$ formulation}
The single-particle Hamiltonian is
\begin{align}
H & = \frac{{\bf p}^2}{2m_0} + V({\bf r}) + \frac{\hbar}{4m_0^2 c^2}\left( {\bf \nabla}V\times {\bf p}\right)\cdot {\bf \sigma} + e{\bm{ \mathcal{E}}}\cdot {\bf r} \nonumber \\
 & + g_0 \mu_B {\bf \sigma} \cdot {\bf B} + H_\text{strain},
\end{align}
where $e{\bm{\mathcal E}}\cdot {\bf r}$, $g_0 \mu_B {\bf \sigma} \cdot {\bf B}$, and $H_\text{strain}$ are Stark, Zeeman, and strain terms, respectively. The strain Hamiltonian is unspecified but we include any term allowed by symmetry. We note in passing that Landau levels (orbital magnetic field effects) are invoked by minimal substitution, i.e.,
\begin{align}
 {\bf p} \rightarrow {\bf p} + e{\bf A}.
\end{align}

\begin{table}
\begin{center}
\subfloat[][]{
\begin{tabular}{c|c c c c c c |c }
$D_{3h}$ & $E$ & $\sigma_h$ & $2C_3$ & $2S_3$  & $3C_2$ & $3\sigma_v$ &  \\ \hline
$\Gamma_1$ & 1 & 1 & 1 & 1 & 1 & 1 & $q_{||}^2,\Re[q_+^3]$\\
$\Gamma_2$ & 1 & 1 & 1 & 1 & $-1$ & $-1$ & $\Im[q_+^3]$\\
$\Gamma_3$ & 1 & $-1$ & 1 & $-1$ & 1 & $-1$ & \\
$\Gamma_4$ & 1 & $-1$ & 1 & $-1$ & $-1$ & 1 & \\
$\Gamma_5$ & 2 & $-2$ & $-1$ & 1 & 0 & 0 &  \\
$\Gamma_6$ & 2 & 2 & $-1$ & $-1$ & 0 & 0 & $\{q_+,q_-\},\{q_-^2,q_+^2\}$ \\
\hline
\end{tabular}
}\\
\subfloat[][]{
\begin{tabular}{c|c c c c c c |c }
$C_{3h}$ & $E$ & $S_3^{-1}$ & $C_3$ & $\sigma_h$ & $C_3^{-1}$ & $S_3$ \\ \hline
$\Gamma_1$ & 1 & 1 & 1 & 1 & 1 & 1 & $q_{||}^2,q_+^3,q_-^3$ \\
$\Gamma_2$ & 1 & $-\omega^2$ & $\omega^4$ & 1 & $-\omega^2$ & $\omega^4$ & $q_+,q_-^2$ \\
$\Gamma_3$ & 1 & $\omega^4$ & $-\omega^2$ & 1 & $\omega^4$ & $-\omega^2$ & $q_-,q_+^2$\\
$\Gamma_4$ & 1 & $-1$ & 1 & $-1$ & 1 & $-1$ & \\
$\Gamma_5$ & 1 & $\omega^2$ & $\omega^4$ & $-1$ & $-\omega^2$ & $-\omega^4$ \\
$\Gamma_6$ & 1 & $-\omega^4$ & $-\omega^2$ & $-1$ & $\omega^4$ & $\omega^2$   \\
\hline
\end{tabular}
}
\caption{Character tables. $\omega = e^{i\pi/6}$. From Ref. \onlinecite{koster}.}
\label{chartables}
\end{center}
\end{table}

\begin{table}
\begin{center}
\subfloat[][]{
\begin{tabular}{c|c c c c c c |c }
$D_{3h}$ & $\Gamma_1$ & $\Gamma_2$ & $\Gamma_3$ & $\Gamma_4$  & $\Gamma_5$ & $\Gamma_6$  \\ \hline
$\Gamma_1$ & $\Gamma_1$ & $\Gamma_2$ & $\Gamma_3$ & $\Gamma_4$ & $\Gamma_5$ & $\Gamma_6$  \\
$\Gamma_2$ & $\Gamma_2$ & $\Gamma_1$ & $\Gamma_4$ & $\Gamma_3$ & $\Gamma_5$ & $\Gamma_6$ \\
$\Gamma_3$ & $\Gamma_3$ & $\Gamma_4$ & $\Gamma_1$ & $\Gamma_2$ & $\Gamma_6$ & $\Gamma_5$ \\
$\Gamma_4$ & $\Gamma_4$ & $\Gamma_3$ & $\Gamma_2$ & $\Gamma_1$ & $\Gamma_6$ & $\Gamma_5$ \\
$\Gamma_5$ & $\Gamma_5$ & $\Gamma_5$ & $\Gamma_6$ & $\Gamma_6$ & $\Gamma_1 \oplus \Gamma_2 \oplus \Gamma_6$ & 
$\Gamma_3 \oplus \Gamma_4 \oplus \Gamma_5$ \\
$\Gamma_6$ & $\Gamma_6$ & $\Gamma_6$ & $\Gamma_5$ & $\Gamma_5$ & $\Gamma_3 \oplus \Gamma_4 \oplus \Gamma_5$ & $\Gamma_1 \oplus \Gamma_2 \oplus \Gamma_6$ \\
\hline
\end{tabular}
}\\
\subfloat[][]{
\begin{tabular}{c|c c c c c c |c }
$C_{3h}$ & $\Gamma_1$ & $\Gamma_2$ & $\Gamma_3$ & $\Gamma_4$  & $\Gamma_5$ & $\Gamma_6$  \\ \hline
$\Gamma_1$ & $\Gamma_1$ & $\Gamma_2$ & $\Gamma_3$ & $\Gamma_4$ & $\Gamma_5$ & $\Gamma_6$  \\
$\Gamma_2$ & $\Gamma_2$ & $\Gamma_3$ & $\Gamma_1$ & $\Gamma_5$ & $\Gamma_6$ & $\Gamma_4$ \\
$\Gamma_3$ & $\Gamma_3$ & $\Gamma_1$ & $\Gamma_2$ & $\Gamma_6$ & $\Gamma_4$ & $\Gamma_5$ \\
$\Gamma_4$ & $\Gamma_4$ & $\Gamma_5$ & $\Gamma_6$ & $\Gamma_1$ & $\Gamma_2$ & $\Gamma_3$ \\
$\Gamma_5$ & $\Gamma_5$ & $\Gamma_6$ & $\Gamma_4$ & $\Gamma_2$ & $\Gamma_3$ & $\Gamma_1$ \\
$\Gamma_6$ & $\Gamma_6$ & $\Gamma_4$ & $\Gamma_5$ & $\Gamma_3$ & $\Gamma_1$ & $\Gamma_2$ \\
\hline
\end{tabular}
}
\caption{Product tables of irreducible representations. From Ref. \onlinecite{koster}.}
\label{charprodtables}
\end{center}
\end{table}
\subsection{$\mathbf K$ point}
From DFT calculations we know that the fundamental band gap occurs at the $K$ point, and the highest valence and lowest conduction bands both consist of a single $p_z$ orbital doubly degenerate due to spin. The group of the wave vector at $\mathbf K$ is $C_{3h}$~\cite{Dresselhaus}, which has only one-dimensional representations as shown in Table \ref{chartables}.  We define $\mathbf q$ as the wave vector relative to the $\mathbf K$ point $\mathbf q = \mathbf k - \mathbf K$, and $q_\pm = q_x \pm i q_y$ transform according to $\Gamma_2$ and $\Gamma_3$ respectively. The $p_z$ orbitals transform according to the $\Gamma_4$ representation and since $\Gamma_4 \otimes \Gamma_4 = \Gamma_1$ there is no direct $k\cdot p$ coupling between the valence and conduction bands. We consider perturbation theory to fourth order in $\mathbf q$. Only invariant terms in $\mathbf q$ are allowed for both diagonal and off-diagonal terms since valence and conduction bands have identical representations. The allowed terms up to fourth order in $\mathbf q$ are $q_{||}^2,q_+^3,q_-^3,q_{||}^4$. However, the combination of reflection symmetry $\sigma_v$ in the $yz$ plane and time-reversal (TR) symmetry limits the possible third-order diagonal terms. Since we have that $\sigma_v: \mathbf K + (q_x,q_y) \rightarrow -\mathbf K - (q_x,-q_y)$ and $TR:-\mathbf K - (q_x,-q_y) \rightarrow \mathbf K + (q_x,-q_y)$, the dispersion around $\mathbf K$ must be even in $q_y$. The only allowed third-order diagonal term is therefore $\Re[q_+^3]=q_x^3 -3q_xq_y^2$. The $k\cdot p$ Hamiltonian in the basis $\ket{p_z},\ket{p_z'}$ is given by:
\begin{align}
    &H^\mathbf{K} = \\
    &\begin{pmatrix}
    E^{\mathbf{K} c} + a q_{||}^2 +  b \Re[q_+^3] + c q_{||}^4 & d q_{||}^2  \\
    d^* q_{||}^2  & E^{\mathbf{K} v} + a' q_{||}^2 +  b' \Re[q_+^3] + c' q_{||}^4
    \end{pmatrix}. \nonumber
\end{align}
Only second order terms are included in the off-diagonal entries, since higher-order terms do not contribute up to fourth order in the dispersion. Diagonalizing this Hamiltonian and expanding to fourth order in $\mathbf q$ gives the following dispersions for the valence and conduction bands:
\begin{subequations}
\begin{align}
    E_c(\mathbf q) &= E^{\mathbf{K} c} + a q_{||}^2 +  b \Re[q_+^3] + \bar{c}  q_{||}^4, \\
    E_v(\mathbf q) &= E^{\mathbf{K} v} + a' q_{||}^2 +  b' \Re[q_+^3] + \bar{c}' q_{||}^4,
\label{dispersion_K}
\end{align}
\end{subequations}
where $\bar{c} = \left(c + \frac{|d|^2}{E^{\mathbf{K} c}-E^{\mathbf{K} v}}\right)$ and $\bar{c}'=\left(c' + \frac{|d|^2}{E^{\mathbf{K} v}-E^{\mathbf{K}_c}}\right)$.
Now we consider the effect of spin-orbit coupling. The SOC Hamiltonian is given by:
\begin{align}
    H_{\text{soc}} = \frac{\hbar}{4m_0^2c^2}(\nabla V \times \mathbf p)\cdot \sigma = \frac{\hbar}{4m_0^2c^2}\mathbf N\cdot \sigma.
\end{align}
The operators $N_\pm = N_x \pm i N_y$ transform according to the representations $\Gamma_5$ and $\Gamma_6$, respectively, and we have no nonzero matrix elements between the valence and conduction band. $N_z$ is an invariant leading to the following expression for the spin-orbit Hamiltonian (in the basis $\ket{p_z\uparrow},\ket{p_z'\uparrow},\ket{p_z\downarrow},\ket{p_z'\downarrow}$):
\begin{align}
    H^\mathbf{K}_\text{SOC} = \begin{pmatrix}
    \Delta_1 & \Delta_3 & 0 & 0 \\
    \Delta_3^* & \Delta_2 & 0 & 0 \\
    0 & 0 & -\Delta_1 & -\Delta_3 \\
    0 & 0 & -\Delta_3^* & - \Delta_2
    \end{pmatrix}.
\end{align}
However, from our DFT calculations we find that the effect is negligible.

It is straightforward to extend our analysis to include strain and external electric and magnetic fields. To linear order in strain we have $\epsilon_{||}=\epsilon_{xx} + \epsilon_{yy}$ which is invariant, and $\epsilon_{+}=\epsilon_{xx} -\epsilon_{yy} + 2i\epsilon_{xy}$ ($\epsilon_{-}=\epsilon_{xx} -\epsilon_{yy} - 2i\epsilon_{xy}$) transforming according to $\Gamma_2$ ($\Gamma_3$). Including only linear strain terms we get the Hamiltonian ($\ket{p_z},\ket{p_z'}$):
\begin{align}
H^\mathbf{K}_\text{strain} = \begin{pmatrix} D_1 \epsilon_{||} & D_2 \epsilon_{||} \\
D_2^*\epsilon_{||} & D_3 \epsilon_{||}.
\end{pmatrix}.
\label{strain_K}
\end{align}
Diagonalizing $H^\mathbf{K} + H^\mathbf{K}_\text{strain}$ for $\mathbf q=0$ and expanding to lowest order in $\epsilon_{||}$ gives the energies at the $\mathbf K$ point:
\begin{subequations}
\begin{align}
E^{\mathbf{K} c}(\bm \epsilon) = E^{\mathbf{K} c} + D_1 \epsilon_{||} \\
E^{\mathbf{K} v}(\bm{\epsilon}) = E^{\mathbf{K} v} + D_3 \epsilon_{||}.
\end{align}
\label{K_strain}
\end{subequations}

For the electric field, $\bm{\mathcal E}$, we have the invariants $\mathcal E_z^2$ and $\mathcal E_{||}^2 = \mathcal E_x^2 + \mathcal E_y^2$, thus in the basis ($\ket{p_z},\ket{p_z'}$):
\begin{align}
H^\mathbf{K}_\text{E-field} = \begin{pmatrix} A_1 \mathcal E_z^2 + A_2 \mathcal E_{||}^2 & A_3 \mathcal E_z^2 + A_4 \mathcal E_{||}^2 \\
A_3^* \mathcal E_z^2 + A_4^* \mathcal E_{||}^2 & A_5 \mathcal E_z^2 + A_6 \mathcal E_{||}^2
\end{pmatrix}.
\end{align}
Again we expand the energies to lowest order in the electric field giving:
\begin{subequations}
\begin{align}
E^{\mathbf{K} c}(\bm{\mathcal E}) = E^{\mathbf{K} c} +  A_1 \mathcal E_z^2 + A_2 \mathcal E_{||}^2\\
E^{\mathbf{K} v}(\bm{\mathcal E}) = E^{\mathbf{K} v} + A_5 \mathcal E_z^2 + A_6 \mathcal E_{||}^2.
\end{align}
\label{K_efield}
\end{subequations}

Finally, in the presence of a magnetic field we have both a Zeeman and a Landau contribution ($\ket{p_z\uparrow},\ket{p_z'\uparrow},\ket{p_z\downarrow},\ket{p_z'\downarrow}$):
\begin{align}
    H^\mathbf{K}_\text{Zeeman} =  \begin{pmatrix} g_1 \mu_B B_z & 0 & g_{1}\mu_B B_- & 0\\
    0 & g_2\mu_B  B_z& 0 &  g_2 \mu_B B_- \\
    g_1 \mu_B B_+ & 0 & -g_1 \mu_B B_z & 0 \\
    0 & g_2 \mu_B B_+ & 0 & -g_2 \mu_B B_z
    \end{pmatrix},
\end{align}

\begin{align}
H^\mathbf{K}_\text{Landau} = \begin{pmatrix}
\alpha_1 B_z & \alpha_2 B_z & 0 & 0 \\
\alpha_2^* B_z & \alpha_3 B_z & 0 & 0 \\
0  & 0 &  \alpha_1 B_z & \alpha_2 B_z \\
0 & 0 & \alpha_2^* B_z & \alpha_3 B_z
\end{pmatrix}.
\end{align}

\subsection{$\mathbf \Gamma$ Point}
We now consider the valence band at the $\mathbf \Gamma$ point. DFT calculations reveal that the valence band is spanned by a $p_\pm = p_x \pm i p_y$ pair of orbitals. The group of the wave vector at the $\Gamma$ point is $D_{3h}$, and the $p_\pm$ orbitals transform according to the $\Gamma_6$ representation, and are therefore degenerate at the $\mathbf \Gamma$ point without spin-orbit coupling. First we derive the $k \cdot p$ Hamiltonian in the absence of spin-orbit coupling. There are no direct $k\cdot p$ terms between the two bands since:
\begin{align}
    \bra{p_+} \mathbf{p}\ket{p_-} = \bra{p_+} \tfrac{m_0}{i\hbar}[H_0,\mathbf r]\ket{p_-} = 0,
\end{align}
where we used $E_0(p_+) = E_0(p_-)$. Next, we consider second-order L\"{o}wdin perturbation theory including interaction terms with $s$ bands. Along the diagonal we get:
\begin{align}
\frac{\hbar^2k_{||}^2}{2m_0} + \frac{\hbar^2}{m_0^2}\sum_s \frac{\bra{p_\pm} \mathbf k \cdot \mathbf p \ket{s}\bra{s} \mathbf k \cdot \mathbf p \ket{p_{\pm}}}{E_s - E_p} = a^{\mathbf{\Gamma} v} k_{||}^2,
\end{align}
defining the constant $a^{\mathbf{\Gamma} v}$. Similarly we have the off diagonal terms:
\begin{align}
\frac{\hbar^2}{m_0^2}\sum_s \frac{\bra{p_+} \mathbf k \cdot \mathbf p \ket{s}\bra{s} \mathbf k \cdot \mathbf p \ket{p_-}}{E_s - E_p} =  c^{\mathbf{\Gamma} v} k_{-}^2.
\end{align}
Thus without spin-orbit coupling we have to second order in $\mathbf k$ the Hamiltonian ($\ket{p_+},\ket{p_-}$):
\begin{align}
    H^{\mathbf{\Gamma} v} = E^{\mathbf{\Gamma} v} + \begin{pmatrix}
    a^{\mathbf{\Gamma} v} k_{||}^2 & c^{\mathbf{\Gamma} v} k_-^2 \\
    c^{\mathbf{\Gamma} v *} k_+^2 & a^{\mathbf{\Gamma} v} k_{||}^2
    \end{pmatrix},
\end{align}
giving the dispersion:
\begin{align}
    E^{\mathbf{\Gamma} v}_{\pm} (\mathbf k) = E^{\mathbf{\Gamma} v} + (a^{\mathbf{\Gamma} v} \pm |c^{\mathbf{\Gamma} v}|) k_{||}^2,
\label{Gamma val_nosoc}
\end{align}
where each subband is doubly degenerate due to spin. Now, consider the effect of spin-orbit coupling. 
The operators $N_\pm = N_x \pm i N_y$ transform according to the representation $\Gamma_5$, while $N_z$ transforms according to $\Gamma_2$. Since $\Gamma_6\otimes\Gamma_5\otimes\Gamma_6 =\Gamma_5 \oplus \Gamma_5 \oplus \Gamma_3 \oplus \Gamma_4 \oplus \Gamma_5$ does not contain the invariant representation there are no non-zero matrix elements between the $p_\pm$ states and $N_\pm$.. But the diagonal matrix elements are allowed since $N_z$ belongs to $\Gamma_2$ and $\Gamma_6 \otimes \Gamma_2 \otimes \Gamma_6=\Gamma_1 \oplus \Gamma_2 \oplus \Gamma_6$ contains the invariant representation:
\begin{align}
    \bra{p_+}N_z\ket{p_+} = -\bra{p_-}N_z\ket{p_-} = \Delta.
\end{align}
Including the SOC Hamiltonian using the basis $\ket{p_+ \uparrow},\ket{p_- \uparrow}, \ket{p_+ \downarrow},\ket{p_- \downarrow}$:
\begin{align}
    H^{\mathbf{\Gamma} v}_{SOC} &= 
    \begin{pmatrix}
    -\Delta  & 0 & 0 & 0 \\
    0 & \Delta & 0 & 0 \\
    0 & 0 & \Delta & 0 \\
    0 & 0 & 0 & -\Delta 
    \end{pmatrix} ,
\end{align}
we find that the degeneracy is broken from 4 to 2 at $\mathbf \Gamma$, but we still have two doubly degenerate bands, with the dispersion:
\begin{align}
    E^{\mathbf{\Gamma} v}_\pm(\mathbf k) = E^{\mathbf{\Gamma} v} +  a^{\mathbf{\Gamma} v} k_{||}^2 \pm \sqrt{\Delta^2 + |c^{\mathbf{\Gamma} v}|^2 k_{||}^4}.
\label{Gamma val_soc}
\end{align}

The effect of strain on the valence band is given by the Hamiltonian ($\ket{p_+},\ket{p_-}$)
\begin{align}
    H^{\mathbf{\Gamma} v}_\text{strain} = \begin{pmatrix} D^{\mathbf{\Gamma} v}_1 \epsilon_{||}  & D^{\mathbf{\Gamma} v}_2 \epsilon_+ \\
    {D^{\mathbf{\Gamma} v}_2}^*\epsilon_-  & D^{\mathbf{\Gamma} v}_1 \epsilon_{||} 
    \end{pmatrix}.
\end{align}
Here the diagonal terms are equal  because $\ket{p_\pm}$ are two basis functions for a two-dimensional representation in contrast to $\ket{p_z}$ and $\ket{p_z'}$  having identical, but different representations. Diagonalizing the total Hamiltonian $H^{\mathbf{\Gamma} v}$ and $ H^{\mathbf{\Gamma} v}_\text{strain}$ we get the eigenenergies:
\begin{align}
E^{\mathbf{\Gamma} v}_\pm(\bm \epsilon) =  E^{\mathbf{\Gamma} v} + D^{\mathbf{\Gamma} v}_1 \epsilon_{||}  \pm | D_2^{\mathbf{\Gamma} v}|\sqrt{(\epsilon_{xx}-\epsilon_{yy})^2 + 4\epsilon_{xy}^2},
\label{strain_Gamma val}
\end{align}
showing that for biaxial strain $\epsilon_{xx}=\epsilon_{yy}$ the two states remain degenerate but for shear strain or non-biaxial normal strain, the degeneracy is broken (without incorporating SOC).

For the electric field we get the Hamiltonian ($\ket{p_+},\ket{p_-}$)
\begin{align}
H^{\mathbf{\Gamma} v}_\text{E-field} = \begin{pmatrix} A^{\mathbf{\Gamma} v}_1 \mathcal E_z^2 + A^{\mathbf{\Gamma} v}_2 \mathcal E_{||}^2 & A^{\mathbf{\Gamma} v}_3 \mathcal E_{+}^2  + A^{\mathbf{\Gamma} v}_4 \mathcal E_- \mathcal E_z \\
A^{\mathbf{\Gamma} v *}_3  \mathcal E_-^2 + A^{\mathbf{\Gamma} v *}_4 \mathcal E_+ \mathcal E_z & A^{\mathbf{\Gamma} v}_1 \mathcal E_z^2 + A^{\mathbf{\Gamma} v}_2 \mathcal E_{||}^2
\end{pmatrix}.
\end{align}
Diagonalizing gives the energies at the $\mathbf \Gamma$ point (without SOC):
\begin{align}
E_\pm^{\mathbf{\Gamma} v}(\bm{\mathcal E}) = E^{\mathbf{\Gamma} v} + A^{\mathbf{\Gamma} v}_1 \mathcal E_z^2 + A^{\mathbf{\Gamma} v}_2 \mathcal E_{||}^2 \pm \left|A^{\mathbf{\Gamma} v}_3 \mathcal E_{+}^2  + A^{\mathbf{\Gamma} v}_4 \mathcal E_- \mathcal E_z\right|.
\end{align}
We see that an in-plane electric field can split the $\ket{p_\pm}$ bands, whereas a perpendicular electric field shifts the two bands equally.

The Zeeman effect is described by the following Hamiltonian
(in the basis states $\ket{p_+}\uparrow,\ket{p_-}\uparrow,\ket{p_+}\downarrow,\ket{p_-}\downarrow$):
\begin{align}
    H^{\mathbf{\Gamma} v}_\text{Zeeman} =  \begin{pmatrix} g^{\mathbf{\Gamma} v}_1 \mu_B B_z & 0 & g^{\mathbf{\Gamma} v}_{1}\mu_B B_- & 0\\
    0 & g^{\mathbf{\Gamma} v}_1\mu_B  B_z& 0 &  g^{\mathbf{\Gamma} v}_1 \mu_B B_- \\
    g^{\mathbf{\Gamma} v}_1 \mu_B B_+ & 0 & -g^{\mathbf{\Gamma} v}_1 \mu_B B_z & 0 \\
    0 & g^{\mathbf{\Gamma} v}_1 \mu_B B_+ & 0 & -g^{\mathbf{\Gamma} v}_1 \mu_B B_z
    \end{pmatrix}.
\end{align}

Due to $\sigma_v$ reflection symmetry in the $yz$ plane, the Landau effect vanishes between the states 
$\ket{p_+},\ket{p_-}$.

DFT calculations reveal that the conduction band at the $\mathbf \Gamma$ point comprises two $s$ bands ($\ket{s},\ket{s'}$) described by the Hamiltonian
\begin{align}
    H^{\mathbf{\Gamma} c} = \begin{pmatrix} E^{\mathbf{\Gamma} c}_1 + a^{\mathbf{\Gamma} c}_1 k_{||}^2 & c^{\mathbf{\Gamma} c} k_{||}^2 \\
   c^{\mathbf{\Gamma} c *} k_{||}^2 & E^{\mathbf{\Gamma} c}_{2} + a^{\mathbf{\Gamma} c}_2 k_{||}^2 .
    \end{pmatrix}
\end{align}
Diagonalizing the Hamiltonian and expanding to lowest order in $k$ leads to the dispersion:
\begin{align}
E^{\mathbf{\Gamma} c}_i =  E^{\mathbf{\Gamma} c}_i + a^{\mathbf{\Gamma} c}_i k_{||}^2,
\label{Gamma val_efield}
\end{align}
where $i\in\{1,2\}$ denotes the two subbands. Since $N_\pm$ transform according to $\Gamma_5$ and $N_z$ according to $\Gamma_2$ there are no spin-orbit interactions between the two conduction band states. The effect of strain on the conduction band is given by the Hamiltonian ($\ket{s},\ket{s'}$)
\begin{align}
H^{\mathbf{\Gamma} c}_\text{strain} = \begin{pmatrix} D^{\mathbf{\Gamma} c}_1 \epsilon_{||} + D^{\mathbf{\Gamma} c}_2 \epsilon_{+}\epsilon_- &   D^{\mathbf{\Gamma} c}_3 \epsilon_{||} +  D^{\mathbf{\Gamma} c}_4 \epsilon_{+}\epsilon_- \\
 D^{\mathbf{\Gamma} c *}_3 \epsilon_{||} +  D^{\mathbf{\Gamma} c *}_4 \epsilon_{+}\epsilon_-  & D^{\mathbf{\Gamma} c}_5 \epsilon_{||} + D^{\mathbf{\Gamma} c}_6 \epsilon_{+}\epsilon_-
\end{pmatrix}.
\label{Gamma con}
\end{align}
Here we have included a second-order term proportional to $\epsilon_+\epsilon_-=(\epsilon_{xx}-\epsilon_{yy})^2 + 4\epsilon_{xy}^2$, since there is no first-order term and from DFT we see an effect of this within reasonable strain magnitudes. The lowest-order contributions to the energies at the $\mathbf \Gamma$ point are then given by:
\begin{subequations}
\begin{align}
E_1^{\mathbf{\Gamma} c}(\bm \epsilon) &= E_1^{\mathbf{\Gamma} c} +  D^{\mathbf{\Gamma} c}_1 \epsilon_{||} + D^{\mathbf{\Gamma} c}_2 \epsilon_{+}\epsilon_- \\
E_2^{\mathbf{\Gamma} c}(\bm \epsilon) &= E_2^{\mathbf{\Gamma} c} +  D^{\mathbf{\Gamma} c}_5 \epsilon_{||} + D^{\mathbf{\Gamma} c}_6 \epsilon_{+}\epsilon_- .
\end{align}
\end{subequations}

The effect of an electric field is described by the Hamiltonian
\begin{align}
H^{\mathbf{\Gamma} c}_\text{E-field} = \begin{pmatrix} A^{\mathbf{\Gamma} c}_1 \mathcal E_z^2 + A^{\mathbf{\Gamma} c}_2 \mathcal E_{||}^2 & A^{\mathbf{\Gamma} c}_3 \mathcal E_z^2 + A^{\mathbf{\Gamma} c}_4 \mathcal E_{||}^2 \\
A^{\mathbf{\Gamma} c *}_3 \mathcal E_z^2 + A^{\mathbf{\Gamma} c  *}_4 \mathcal E_{||}^2 & A^{\mathbf{\Gamma} c}_5 \mathcal E_z^2 + A^{\mathbf{\Gamma} c}_6 \mathcal E_{||}^2
\end{pmatrix}.
\end{align}
Again, we diagonalize and expand to lowest order in the electric field, giving:
\begin{subequations}
\begin{align}
E_1^{\mathbf{\Gamma} c}(\bm{\mathcal E}) &= E_1^{\mathbf{\Gamma} c} +   A^{\mathbf{\Gamma} c}_1 \mathcal E_z^2 + A^{\mathbf{\Gamma} c}_2 \mathcal E_{||}^2 \label{Gamma con1_efied_a} \\
E_2^{\mathbf{\Gamma} c}(\bm{\mathcal E}) &= E_2^{\mathbf{\Gamma} c} +  A^{\mathbf{\Gamma} c}_5 \mathcal E_z^2 + A^{\mathbf{\Gamma} c}_6 \mathcal E_{||}^2 \label{Gamma con1_efied_b} .
\end{align}
\end{subequations}

Finally for a magnetic field we have both a Zeeman and a Landau contribution ($\ket{s\uparrow},\ket{s' \uparrow},\ket{s\downarrow},\ket{s'\downarrow}$):
\begin{align}
    H^{\mathbf{\Gamma} c}_\text{Zeeman} =  \begin{pmatrix} g^{\mathbf{\Gamma} c}_1 \mu_B B_z & 0 & g^{\mathbf{\Gamma} c}_{1}\mu_B B_- & 0\\
    0 & g^{\mathbf{\Gamma} c}_2\mu_B  B_z& 0 &  g^{\mathbf{\Gamma} c}_2 \mu_B B_- \\
    g^{\mathbf{\Gamma} c}_1 \mu_B B_+ & 0 & -g^{\mathbf{\Gamma} c}_1 \mu_B B_z & 0 \\
    0 & g^{\mathbf{\Gamma} c}_2 \mu_B B_+ & 0 & -g^{\mathbf{\Gamma} c}_2 \mu_B B_z
    \end{pmatrix},
\end{align}

\begin{align}
H^{\mathbf{\Gamma} c}_\text{Landau} = \begin{pmatrix}
\alpha^{\mathbf{\Gamma} c}_1 B_z & \alpha^{\mathbf{\Gamma} c}_2 B_z & 0 & 0 \\
\alpha^{\mathbf{\Gamma} c *}_2 B_z & \alpha^{\mathbf{\Gamma} c}_3 B_z & 0 & 0 \\
0  & 0 &  \alpha^{\mathbf{\Gamma} c}_1 B_z & \alpha^{\mathbf{\Gamma} c}_2 B_z \\
0 & 0 & \alpha^{\mathbf{\Gamma} c *}_2 B_z & \alpha^{\mathbf{\Gamma} c}_3 B_z
\end{pmatrix}.
\end{align}

The symmetry analysis carried out here relies only on the symmetry of the crystal and the orbital characters of the relevant bands and therefore can be generalized to other materials sharing the same point group $D_{3h}$. In particular, we obtain using DFT that for hexagonal boron nitride (h-BN) the conduction band is a $s$ orbital and the valence band a $p_z$ orbital at the $\mathbf K$ point. At the $\mathbf \Gamma$ point, the lowest conduction bands are $p_z$ and $s$ states while the valence band is a pair of $p_\pm$ orbitals. Hence, the present Hamiltonian models for the valence bands at the $\mathbf \Gamma$ and $\mathbf K$ points can be directly applied to hexagonal boron nitride as well.

\section{Computational method and extraction of parameters}
We determine parameters of the models derived in the previous section by fitting to ab initio calculations. We have performed standard density functional theory (DFT) calculations using the projector augmented wave method as implemented in the GPAW package\cite{GPAW1,GPAW2} with the Perdew-Burke-Ernzerhof approximation of the exchange-correlation energy \cite{PBE}. The selfconsistent calculation used a 16x16 Monkhorst-Pack $k$ point grid and  a $\SI{1200}{\electronvolt}$ cut-off energy for the plane-wave representation of the wave functions. The structure was relaxed using the Broyden-Fletcher-Goldfarb-Shanno algorithm as implemented in ASE \cite{ase}, resulting in a lattice constant of \SI{3.39}{\angstrom}. We use a vacuum layer of $\SI{20}{\angstrom}$ between adjacent layers to avoid interaction between supercells.  The SOC Hamiltonian was diagonalized in a basis of wave functions from scalar-relativistic calculations as documented in Ref.~\onlinecite{SOC}.

\begin{figure}
    \centering
\subfloat[][]{\includegraphics[width=0.7\columnwidth]{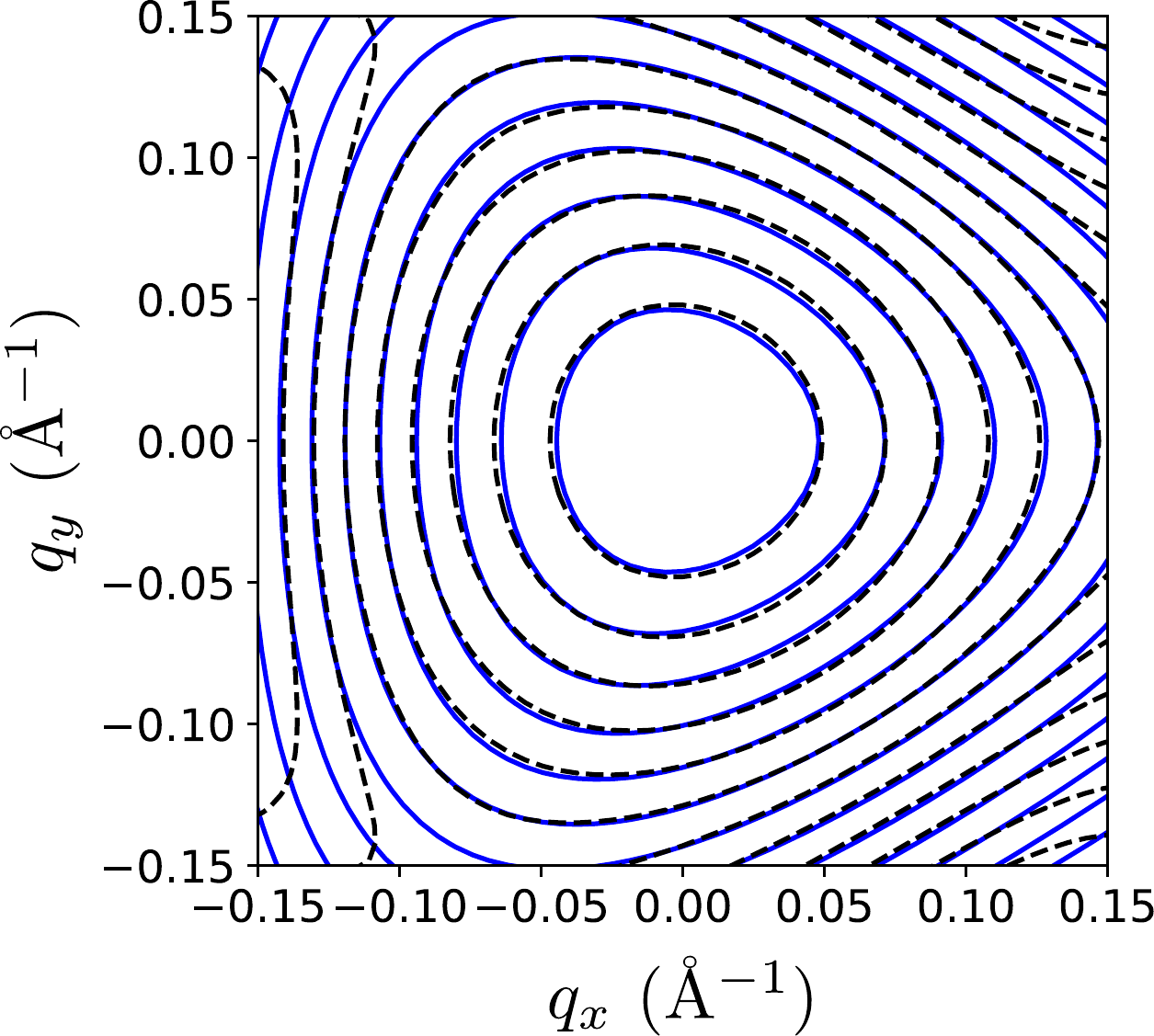}}\\
    \subfloat[][]{\includegraphics[width=0.7\columnwidth]{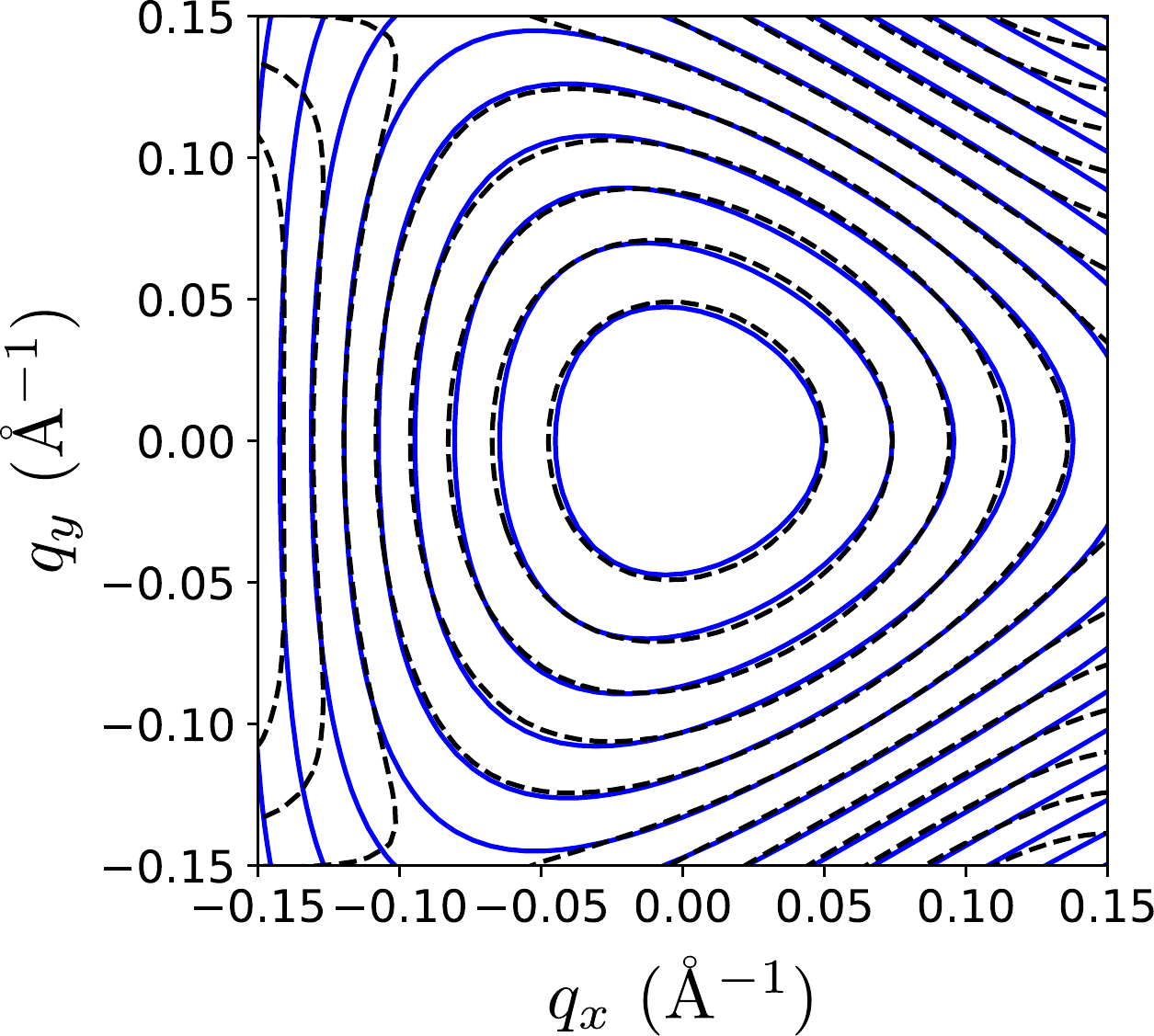}}
    \caption{Contour plot of the dispersion of the (a) valence  and (b) conduction bands around $K$ from DFT (blue solid) and $k\cdot p$ (black dashed), with a spacing between contours of $\SI{0.05}{\electronvolt}$. The trigonal warping is well described by the third-order terms in $\mathbf{q}=\mathbf{k}-\mathbf{K}$, whereas the inclusion of fourth-order terms makes the model valid in a larger area around  $\mathbf K$.}
    \label{fig:K_point}
\end{figure}

To extract the $k\cdot p$ parameters we perform bandstructure calculations on a dense $k$ point grid in the the vicinity of either $\mathbf{K}$ or $\mathbf{\Gamma}$. In Fig.~\ref{fig:K_point}, a contour plot of the valence and conduction bands at the $\mathbf K$ point is shown using both DFT and $k\cdot p$ results from Eq.~\eqref{dispersion_K} with parameters fitted to the DFT energies. The model captures the trigonal warping of the bands described by the third-order terms in $\mathbf q$. We clearly see that the bands are symmetric under $q_y \rightarrow - q_y$ due to the combination of TR and the $\sigma_v$ reflection as explained in the previous section. The $k\cdot p$ model gives very good agreement with DFT energies up to a distance of $\SI{0.1}{\per\angstrom}$ from the $\mathbf K$ point, due to the inclusion of fourth-order terms in $\mathbf{q}$.

\begin{figure}
    \centering
        \subfloat[][]{\includegraphics[width=0.8\columnwidth]{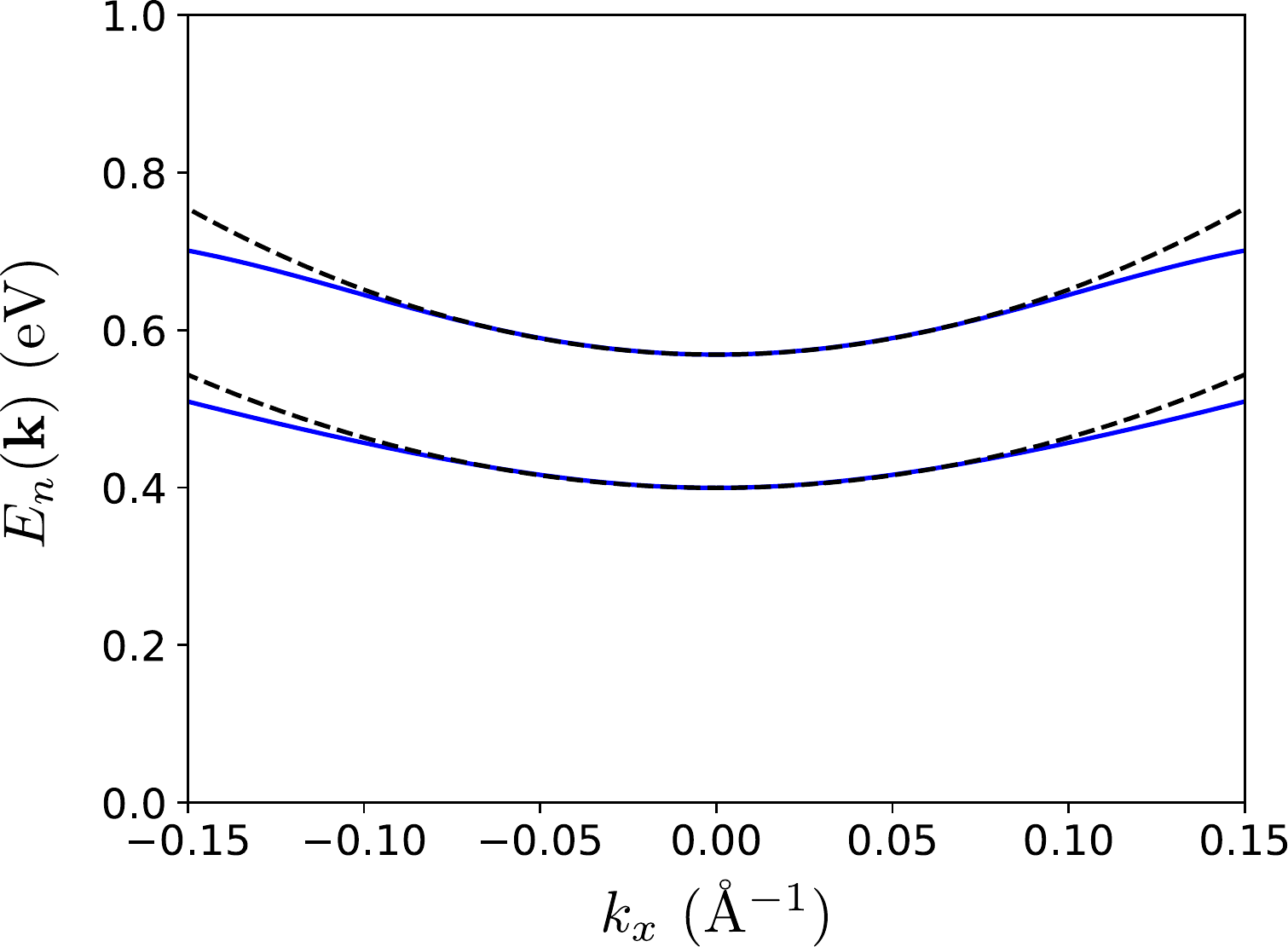} \label{fig:Gamma con}}\\
    \subfloat[][]{\includegraphics[width=0.8\columnwidth]{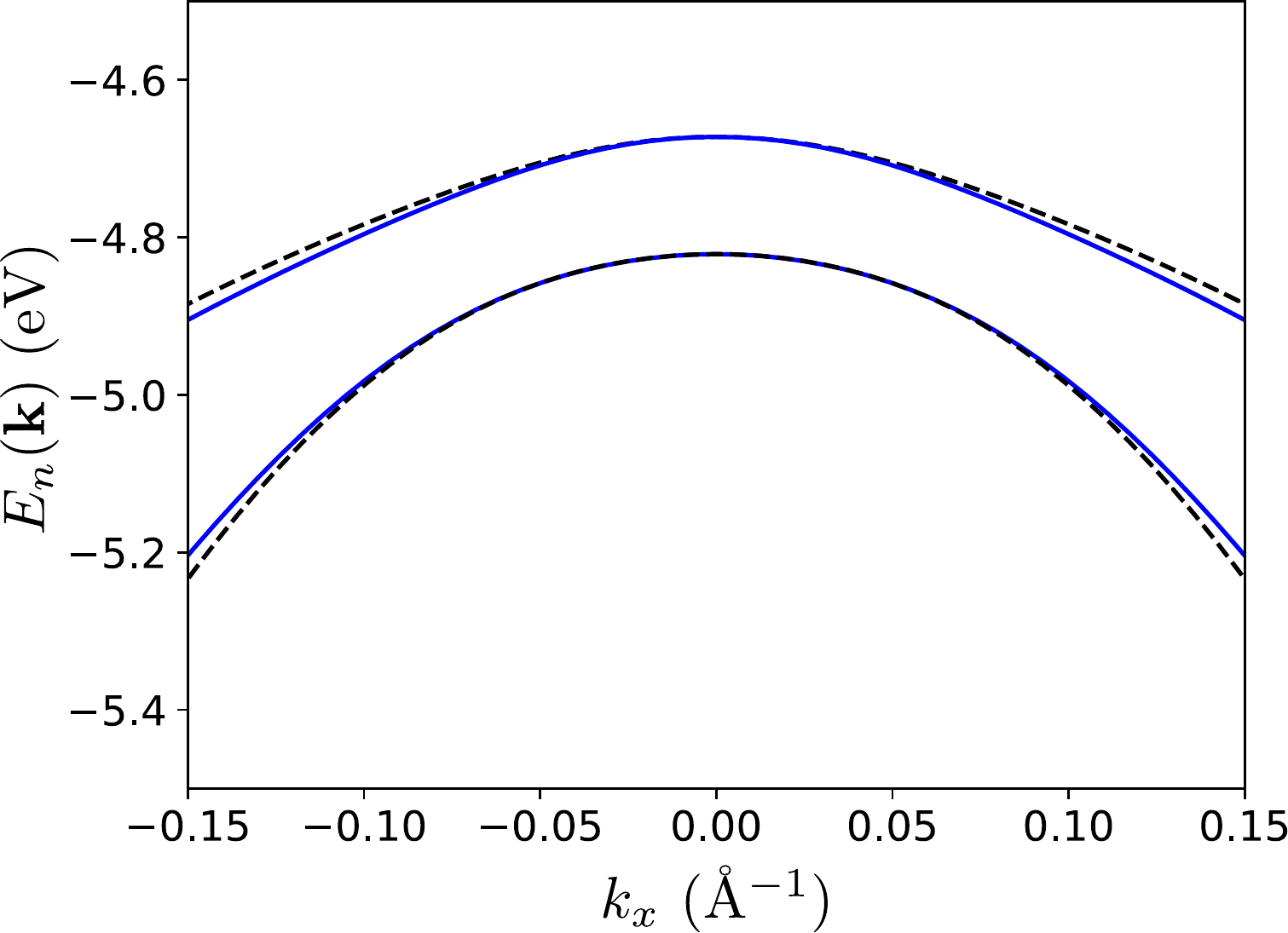}\label{fig:Gamma val}}
    \caption{Dispersion around the $\Gamma$ point from DFT (blue solid) and $k \cdot p$ (black dashed) in the $k_x$ direction including spin-orbit coupling. The coefficients of the $k\cdot p$ terms are obtained from a fit to the DFT result without spin-orbit coupling and the spin-orbit constant is simply obtained from the eigenvalues at $\Gamma$.}
    \label{fig:gamma}
\end{figure}

At the $\mathbf \Gamma$ point, $k\cdot p$ parameters are obtained by fitting Eq.~\eqref{Gamma val_nosoc} to the DFT result without SOC, and the SOC parameter is simply the energy splitting at the $\mathbf \Gamma$ point. The dispersions given by Eq.~\eqref{Gamma val_soc} are in good agreement with the DFT results within $\SI{0.1}{\per\angstrom}$ of $\mathbf \Gamma$ as seen in Fig.~\ref{fig:Gamma val}. For the conduction band only the diagonal terms in Eq.~\eqref{Gamma con} contribute to second order in the dispersion, and we find that keeping only those terms gives a relatively good description of the bands as shown in Fig.~\ref{fig:Gamma con}.

\subsection{Electric field}
Using the external potential module implemented in GPAW we perform calculations of the band structure with a constant electric field in the $z$ direction. In accordance with the symmetry analysis in the previous section we see a parabolic dependence of the eigenvalues on the strength of the electric field. Interestingly, most bands are only slightly changed, but the lowest conduction band at the $\mathbf \Gamma$ point shows high sensitivity to electric fields. At large  electric fields the conduction band minimum changes from the $\mathbf K$ point to the $\mathbf \Gamma$ point at which point BAs becomes an indirect bandgap material. At electric fields larger than \SI{0.75}{\volt\per\angstrom}, the conduction band decreases below the valence band maximum at $\mathbf K$ and we get a transition to a metallic state. In comparison, an electric field induced bandgap of $\SI{0.25}{\electronvolt}$ has been demonstrated experimentally\cite{graphene_zhang} in bilayer graphene subject to an electric displacement field of $\SI{0.3}{\volt\per\angstrom}$. We fit Eqs. \eqref{Gamma con1_efied_a}-\eqref{Gamma con1_efied_b} to the DFT results to obtain the relevant parameters.The second conduction band at $\mathbf \Gamma$ is not fitted since higher-lying bands are shifted down and mixed with this band.

\begin{figure}
    \centering
        \subfloat[][]{\includegraphics[width=0.8\columnwidth]{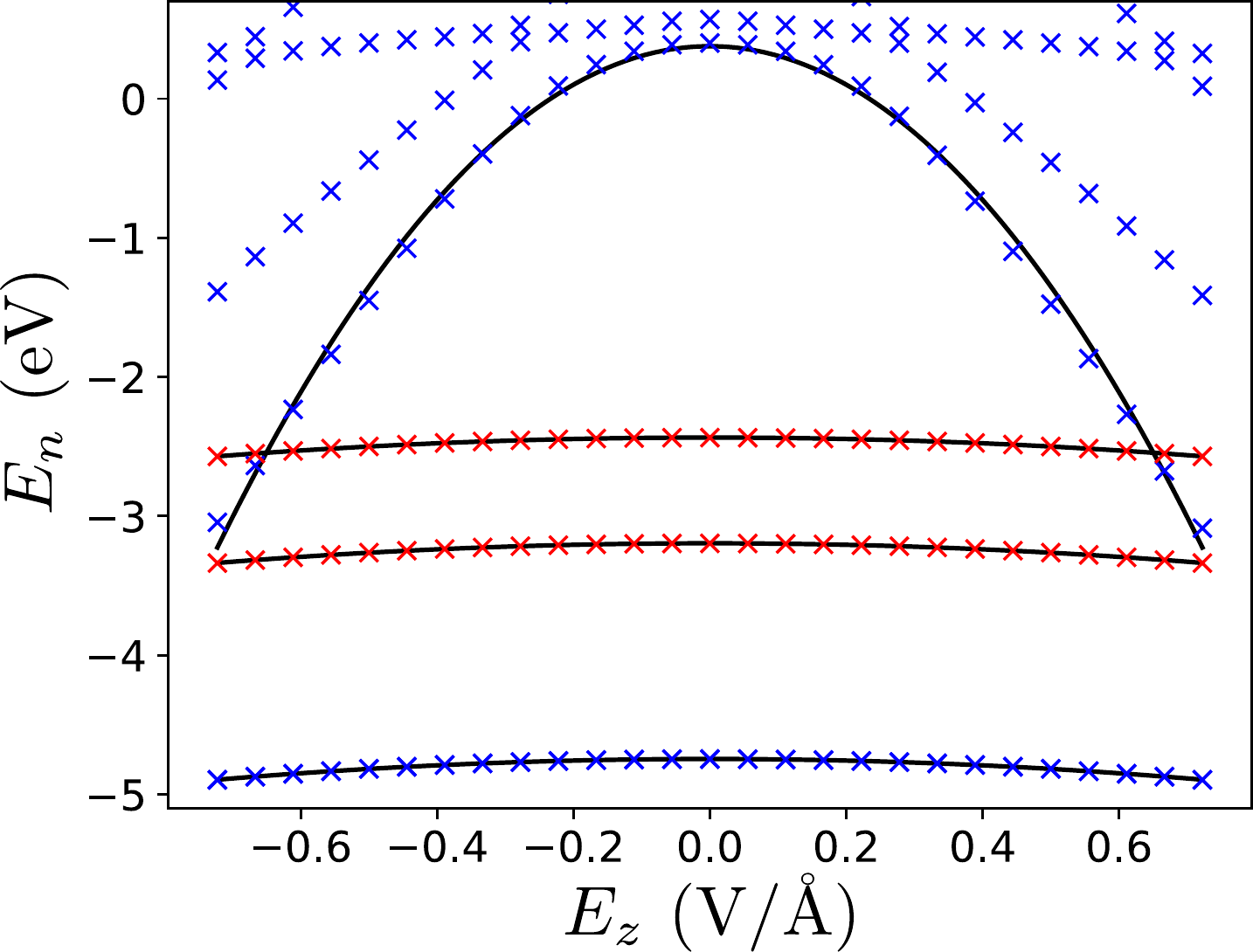}}\\
    \subfloat[][]{\includegraphics[width=0.8\columnwidth]{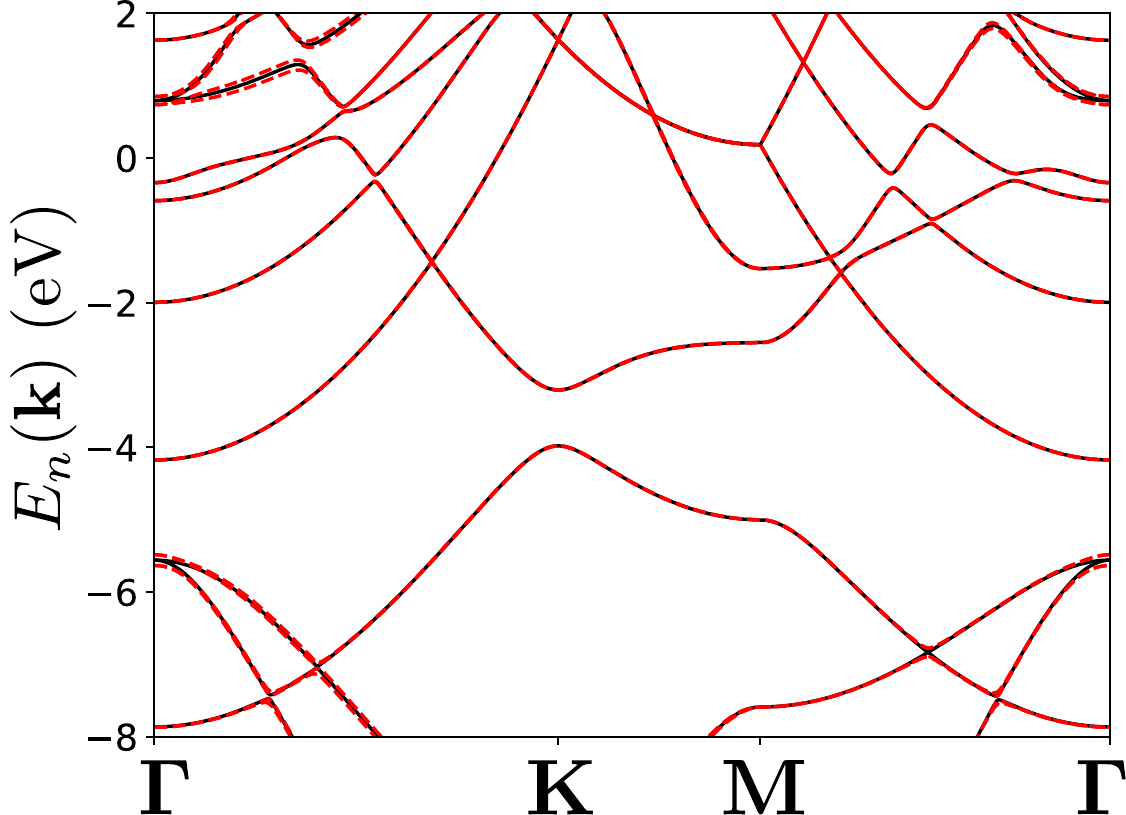}}
    \caption{(a) Eigenenergies at the $\mathbf K$ (red) and $\mathbf \Gamma$ (black) points as a function of electric field in perpendicular to the boron arsenide sheet.  Solid lines are $k\cdot p$ results, crosses are from DFT calculations. (b) DFT bandstructure in the metallic state for an electric field of $\SI{0.78}{\volt\per\angstrom}$ with (red dashed) and without (black solid) SOC.}
    \label{fig:valence_gamma}
\end{figure}

\subsection{Strain}
\begin{figure}
    \centering
        \subfloat[][]{\includegraphics[width=0.8\columnwidth]{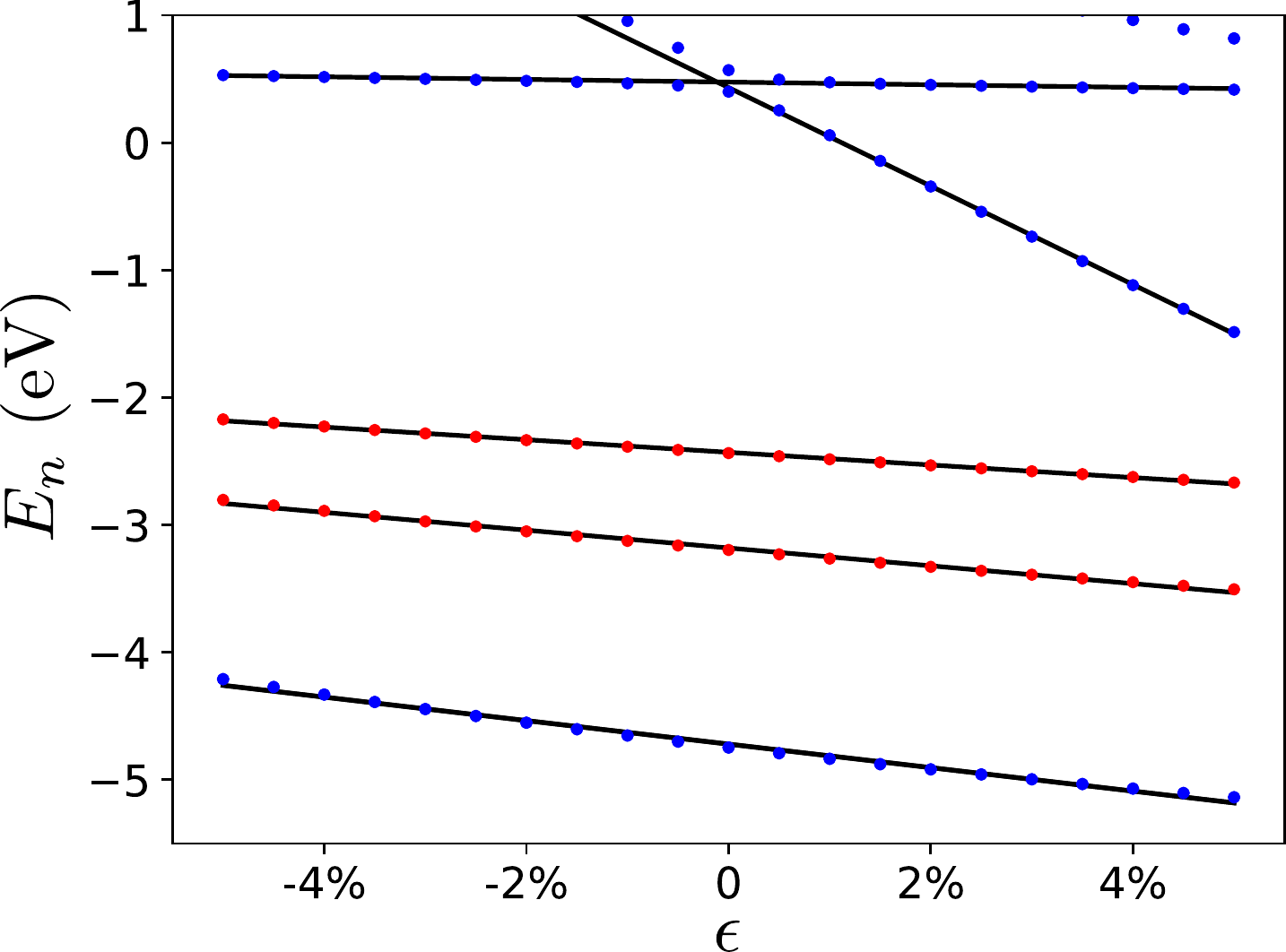}\label{fig:biaxial}}\\
    \subfloat[][]{\includegraphics[width=0.8\columnwidth]{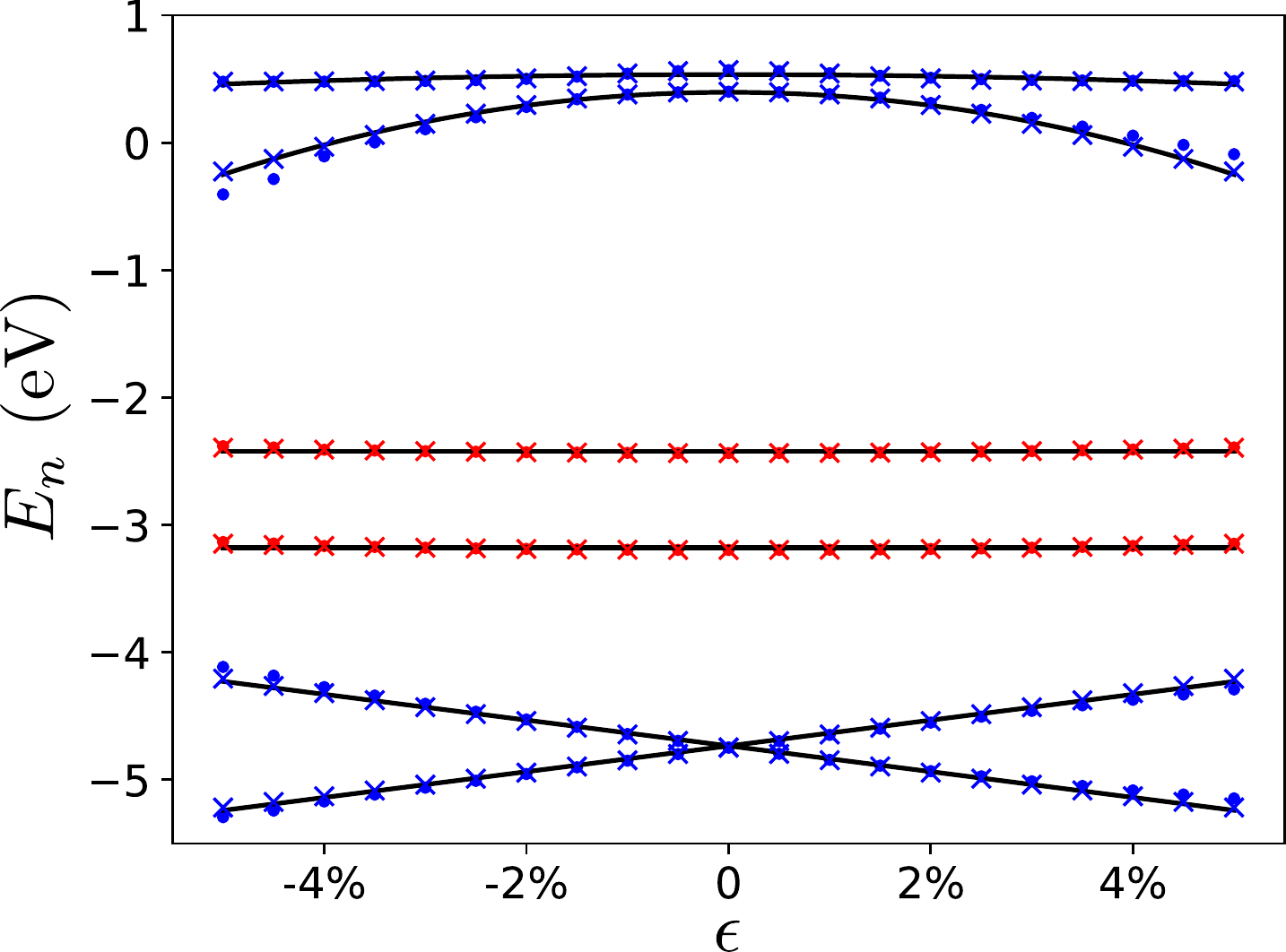}\label{fig:difshear}}
    \caption{(a) Eigenenergies at the $\mathbf K$ (red) and $\mathbf \Gamma$ (black) points as a function of biaxial strain $\epsilon=\epsilon_{xx}=\epsilon_{yy}$ (b) Eigenenergies at the $\mathbf K$ (red) and $\mathbf \Gamma$ (black) as a function of symmetry-breaking normal strain $\epsilon=\epsilon_{xx}=-\epsilon_{yy}$ (dots) and shear strain $\epsilon=\epsilon_{xy}$. Solid lines are $k\cdot p$ results, dots and crosses are DFT results. We see that there is a slight difference between the two cases, which can be explained by including higher-order terms in the symmetry analysis.}
    \label{fig:strain}
\end{figure}

To investigate the effect of strain in the linear approximation we modify the unit cell after the relaxation, $\mathbf a_n' = \mathbf a_n + \bm \epsilon \mathbf a_n$ where $\mathbf \epsilon$ is the strain tensor, $\mathbf a_n$ are the equilibrium lattice vectors, and $\mathbf a_n'$ are the strained lattice vectors. After straining the unit cell we perform a relaxation of the atomic positions while fixing the unit cell fixed to the strained value. There are three independent terms in the in-plane strain, biaxial strain with $\epsilon_{xx}=\epsilon_{yy}$, $\epsilon_{xx}=-\epsilon_{yy}$, and shear strain $\epsilon_{xy}$. Only in the biaxial case is the symmetry of the lattice preserved, and we note that for the band structure calculation the Brillouin zone and the high symmetry points are changed. Hence, we  consider the point in $\mathbf{k}$ space that reduces to $\mathbf K$ when strain goes to zero and we simply denote it $\mathbf K$. We limit our study to the cases where only one of the strain terms is nonzero. At the $\mathbf K$ point we find that the only linear strain terms that occur in the Hamiltonian, Eq.~\eqref{strain_K}, are proportional to $\epsilon_{||}$. By fitting Eq.~\eqref{K_strain} to the DFT results we determine the parameters $D_1$ and $D_3$ and obtain excellent agreement within a range of $\epsilon_{||} =\pm 10\%$ as seen in Fig.~\ref{fig:biaxial}. In Fig.~\ref{fig:difshear} we observe that the eigenvalues at the $\mathbf K$ point remain constant with both $\epsilon_{xx} -\epsilon_{yy}$ and $\epsilon_{xy}$, showing that the effects of strain is well described by the linear strain terms. For the valence band at the $\Gamma$ point we fit Eq.~\eqref{strain_Gamma val} to the DFT results. The resuls are plotted in Fig.~\ref{fig:strain}, showing good agreement betwen model and DFT calculations. Close inspection shows that higher-order terms contribute giving a slight curvature of the bands, but the energy shifts are small compared to the contribution from the linear terms. The higher-order terms also give a difference between $\epsilon_{xx}- \epsilon_{yy}$ and shear strain $\epsilon_{xy}$ even though they belong to the same representation in the symmetry analysis. This comes from the fact that the shear term is multiplied by the imaginary unit $i$. The conduction bands are well described by the linear term in $\epsilon_{||}$ as well, however, for $\epsilon_{xx}-\epsilon_{yy}$ and $\epsilon_{xy}$ there are no linear terms, but we find a significant parabolic dependence in Fig.~\ref{fig:difshear}. Again, higher-order terms show little effect for large strains yet for most cases the second-order term gives a good approximation.

\begin{table}
\centering
\bgroup
\def\arraystretch{1.5}
\begin{minipage}{0.3\columnwidth}
\subfloat[][]{
\begin{tabular}{|l|r|}
\hline
     $E^{\mathbf{K} c}$  $[\si{\electronvolt}]$ & $-2.44$    \\\hline
     $a$  $[\si{\electronvolt\angstrom^2}]$ & $21.4$  \\\hline
     $b$  $[\si{\electronvolt\angstrom^3}]$ & $-25.0$ \\\hline
     $\bar{c}$  $[\si{\electronvolt\angstrom^4}]$ & $-245$ \\\hline
$E^{\mathbf{K} v}$  $[\si{\electronvolt}]$ & $-3.20$    \\ \hline
     $a'$  $[\si{\electronvolt\angstrom^2}]$ & $-22.2$  \\\hline
     $b'$  $[\si{\electronvolt\angstrom^3}]$ & $20.0$ \\\hline
     $\bar{c}'$  $[\si{\electronvolt\angstrom^4}]$ & $250$ \\\hline
     $E^{\mathbf{\Gamma} v}$  $[\si{\electronvolt}]$ & $-4.75$\\\hline
     $a^{\mathbf \Gamma v}$  $ [\si{\electronvolt\angstrom^2}]$ & $-13.9$ \\\hline
     $c^{\mathbf \Gamma v}$  $ [\si{\electronvolt\angstrom^2}]$ & $7.02$ \\\hline
     $\Delta$  $ [\si{\electronvolt}]$ & $0.074$\\\hline
     $E^{\mathbf{\Gamma} c}_1$  $[\si{\electronvolt}]$ & $0.57$ \\\hline
     $E^{\mathbf{\Gamma} c}_2$  $[\si{\electronvolt}]$ & $0.40$\\\hline
$a^{\mathbf{\Gamma} c}_1$  $ [\si{\electronvolt\angstrom^2}]$ & $8.25$ \\\hline
$a^{\mathbf{\Gamma} c}_2$  $ [\si{\electronvolt\angstrom^2}]$ & $ 6.39$\\\hline
\end{tabular}
}
\end{minipage}
\begin{minipage}{0.6\columnwidth}
\subfloat[][]{
\begin{tabular}{|l|r|}
\hline
   $D_1$  $ [\si{\electronvolt}]$ & $-2.48$  \\\hline
   $D_3$  $ [\si{\electronvolt}]$ &  $-3.50$ \\\hline
   $D^{\mathbf{\Gamma} v}_1$  $ [\si{\electronvolt}]$ & $-4.61$ \\\hline
   $D^{\mathbf{\Gamma} v}_2$  $ [\si{\electronvolt}]$ & $-5.06$ \\\hline
   $D^{\mathbf{\Gamma} c}_1$  $ [\si{\electronvolt}]$ & $-0.517$ \\\hline
   $D^{\mathbf{\Gamma} c}_2$  $ [\si{\electronvolt}]$ & $-7.47$ \\\hline
   $D^{\mathbf{\Gamma} c}_5$  $ [\si{\electronvolt}]$ & $-19.3$ \\\hline
   $D^{\mathbf{\Gamma} c}_6 $  $[\si{\electronvolt}]$ & $-64.5$ \\\hline
\end{tabular}
}\\
\subfloat[][]{
\begin{tabular}{|l|r|}
\hline
$A_1 [\si[per-mode=fraction]{\electronvolt\angstrom^2\per\volt^2}]$ & $-0.259$\\\hline 
$A_5 [\si[per-mode=fraction]{\electronvolt\angstrom^2\per\volt^2}]$ &  $-0.271$\\\hline
$A^{\mathbf{\Gamma} v}_1 [\si[per-mode=fraction]{\electronvolt\angstrom^2\per\volt^2}]$ & $-0.285$ \\\hline
$A^{\mathbf{\Gamma} c}_5 [\si[per-mode=fraction]{\electronvolt\angstrom^2\per\volt^2}]$ & $-6.91$ \\\hline
\end{tabular}
}
\end{minipage}
\egroup
\caption{Extracted $k\cdot p$ model parameters from DFT.}
\label{parametertable}
\end{table}

\section{Conclusion}
Compact and accurate $k\cdot p$ Hamiltonian models for the 2D material hexagonal boron arsenide (h-BAs) are extracted based on group symmetry discussions. Model coefficients are found by comparison with detailed density functional theory calculations. 2D hexagonal boron arsenide is expected to be an important material as it is predicted to have an ultrahigh thermal conductivity and provides a low-bandgap candidate to the well-studied high-bandgap 2D semiconductor material h-BN. We derive $k\cdot p$ Hamlitonians and demonstrate excellent agreement with DFT results in the presence of strain and electric fields at two important points in the Brillouin zone, $\mathbf K$ and $\mathbf \Gamma$. DFT calculations reveal that the bandgap of 2D boron arsenide, located at the $\mathbf K$ point, becomes indirect at sufficiently large electric fields with the conduction band minimum located at the $\mathbf \Gamma$ point. At even larger electric fields or a large strain the material becomes metallic. The influence of an external magnetic field is also discussed. The $k\cdot p$ models derived in this work are useful for practical and efficient device simulations in addition to heterostructures, complicated geometries, or accounting for spatially varying external fields where the periodicity is broken and DFT becomes computationally too demanding.

\section*{Acknowledgments}
MRB and MW gratefully acknowledge financial support from the Danish Council of Independent Research (Natural Sciences) grant no.: DFF-4181-00182. MRB is grateful for financial support from Otto Mønsted Foundation and Augustinus Foundation, as well as Beijing Institute of Nanoenergy and Nanosystems, Chinese Academy of sciences.

\bibliography{references}{}

\begin{thebibliography}{47}%
\makeatletter
\providecommand \@ifxundefined [1]{%
 \@ifx{#1\undefined}
}%
\providecommand \@ifnum [1]{%
 \ifnum #1\expandafter \@firstoftwo
 \else \expandafter \@secondoftwo
 \fi
}%
\providecommand \@ifx [1]{%
 \ifx #1\expandafter \@firstoftwo
 \else \expandafter \@secondoftwo
 \fi
}%
\providecommand \natexlab [1]{#1}%
\providecommand \enquote  [1]{``#1''}%
\providecommand \bibnamefont  [1]{#1}%
\providecommand \bibfnamefont [1]{#1}%
\providecommand \citenamefont [1]{#1}%
\providecommand \href@noop [0]{\@secondoftwo}%
\providecommand \href [0]{\begingroup \@sanitize@url \@href}%
\providecommand \@href[1]{\@@startlink{#1}\@@href}%
\providecommand \@@href[1]{\endgroup#1\@@endlink}%
\providecommand \@sanitize@url [0]{\catcode `\\12\catcode `\$12\catcode
  `\&12\catcode `\#12\catcode `\^12\catcode `\_12\catcode `\%12\relax}%
\providecommand \@@startlink[1]{}%
\providecommand \@@endlink[0]{}%
\providecommand \url  [0]{\begingroup\@sanitize@url \@url }%
\providecommand \@url [1]{\endgroup\@href {#1}{\urlprefix }}%
\providecommand \urlprefix  [0]{URL }%
\providecommand \Eprint [0]{\href }%
\providecommand \doibase [0]{http://dx.doi.org/}%
\providecommand \selectlanguage [0]{\@gobble}%
\providecommand \bibinfo  [0]{\@secondoftwo}%
\providecommand \bibfield  [0]{\@secondoftwo}%
\providecommand \translation [1]{[#1]}%
\providecommand \BibitemOpen [0]{}%
\providecommand \bibitemStop [0]{}%
\providecommand \bibitemNoStop [0]{.\EOS\space}%
\providecommand \EOS [0]{\spacefactor3000\relax}%
\providecommand \BibitemShut  [1]{\csname bibitem#1\endcsname}%
\let\auto@bib@innerbib\@empty
\bibitem [{\citenamefont {Novoselov}\ \emph {et~al.}(2004)\citenamefont
  {Novoselov}, \citenamefont {Geim}, \citenamefont {Morozov}, \citenamefont
  {Jiang}, \citenamefont {Zhang}, \citenamefont {Dubonos}, \citenamefont
  {Grigorieva},\ and\ \citenamefont {Firsov}}]{graphene}%
  \BibitemOpen
  \bibfield  {author} {\bibinfo {author} {\bibfnamefont {K.~S.}\ \bibnamefont
  {Novoselov}}, \bibinfo {author} {\bibfnamefont {A.~K.}\ \bibnamefont {Geim}},
  \bibinfo {author} {\bibfnamefont {S.~V.}\ \bibnamefont {Morozov}}, \bibinfo
  {author} {\bibfnamefont {D.}~\bibnamefont {Jiang}}, \bibinfo {author}
  {\bibfnamefont {Y.}~\bibnamefont {Zhang}}, \bibinfo {author} {\bibfnamefont
  {S.~V.}\ \bibnamefont {Dubonos}}, \bibinfo {author} {\bibfnamefont {I.~V.}\
  \bibnamefont {Grigorieva}}, \ and\ \bibinfo {author} {\bibfnamefont {A.~A.}\
  \bibnamefont {Firsov}},\ }\href {\doibase 10.1126/science.1102896} {\bibfield
   {journal} {\bibinfo  {journal} {Science}\ }\textbf {\bibinfo {volume}
  {306}},\ \bibinfo {pages} {666} (\bibinfo {year} {2004})},\ \Eprint
  {http://arxiv.org/abs/https://science.sciencemag.org/content/306/5696/666.full.pdf}
  {https://science.sciencemag.org/content/306/5696/666.full.pdf} \BibitemShut
  {NoStop}%
\bibitem [{\citenamefont {Novoselov}\ \emph {et~al.}(2005)\citenamefont
  {Novoselov}, \citenamefont {Jiang}, \citenamefont {Schedin}, \citenamefont
  {Booth}, \citenamefont {Khotkevich}, \citenamefont {Morozov},\ and\
  \citenamefont {Geim}}]{graphene2}%
  \BibitemOpen
  \bibfield  {author} {\bibinfo {author} {\bibfnamefont {K.~S.}\ \bibnamefont
  {Novoselov}}, \bibinfo {author} {\bibfnamefont {D.}~\bibnamefont {Jiang}},
  \bibinfo {author} {\bibfnamefont {F.}~\bibnamefont {Schedin}}, \bibinfo
  {author} {\bibfnamefont {T.~J.}\ \bibnamefont {Booth}}, \bibinfo {author}
  {\bibfnamefont {V.~V.}\ \bibnamefont {Khotkevich}}, \bibinfo {author}
  {\bibfnamefont {S.~V.}\ \bibnamefont {Morozov}}, \ and\ \bibinfo {author}
  {\bibfnamefont {A.~K.}\ \bibnamefont {Geim}},\ }\href {\doibase
  10.1073/pnas.0502848102} {\bibfield  {journal} {\bibinfo  {journal}
  {Proceedings of the National Academy of Sciences}\ }\textbf {\bibinfo
  {volume} {102}},\ \bibinfo {pages} {10451} (\bibinfo {year}
  {2005})}\BibitemShut {NoStop}%
\bibitem [{\citenamefont {Hao}\ \emph {et~al.}(2018)\citenamefont {Hao},
  \citenamefont {Marichy},\ and\ \citenamefont
  {Journet}}]{fabrication_review_ald}%
  \BibitemOpen
  \bibfield  {author} {\bibinfo {author} {\bibfnamefont {W.}~\bibnamefont
  {Hao}}, \bibinfo {author} {\bibfnamefont {C.}~\bibnamefont {Marichy}}, \ and\
  \bibinfo {author} {\bibfnamefont {C.}~\bibnamefont {Journet}},\ }\href
  {\doibase 10.1088/2053-1583/aad94f} {\bibfield  {journal} {\bibinfo
  {journal} {2D Materials}\ }\textbf {\bibinfo {volume} {6}},\ \bibinfo {pages}
  {012001} (\bibinfo {year} {2018})}\BibitemShut {NoStop}%
\bibitem [{\citenamefont {Li}\ \emph {et~al.}(2018)\citenamefont {Li},
  \citenamefont {Zhang}, \citenamefont {Guo}, \citenamefont {Huang},
  \citenamefont {Lu}, \citenamefont {Lin}, \citenamefont {Wang}, \citenamefont
  {Du},\ and\ \citenamefont {Gao}}]{fabrication_review_epitaxy}%
  \BibitemOpen
  \bibfield  {author} {\bibinfo {author} {\bibfnamefont {G.}~\bibnamefont
  {Li}}, \bibinfo {author} {\bibfnamefont {Y.-Y.}\ \bibnamefont {Zhang}},
  \bibinfo {author} {\bibfnamefont {H.}~\bibnamefont {Guo}}, \bibinfo {author}
  {\bibfnamefont {L.}~\bibnamefont {Huang}}, \bibinfo {author} {\bibfnamefont
  {H.}~\bibnamefont {Lu}}, \bibinfo {author} {\bibfnamefont {X.}~\bibnamefont
  {Lin}}, \bibinfo {author} {\bibfnamefont {Y.-L.}\ \bibnamefont {Wang}},
  \bibinfo {author} {\bibfnamefont {S.}~\bibnamefont {Du}}, \ and\ \bibinfo
  {author} {\bibfnamefont {H.-J.}\ \bibnamefont {Gao}},\ }\href@noop {}
  {\bibfield  {journal} {\bibinfo  {journal} {Chem. Soc. Rev.}\ }\textbf
  {\bibinfo {volume} {47}},\ \bibinfo {pages} {6073} (\bibinfo {year}
  {2018})}\BibitemShut {NoStop}%
\bibitem [{\citenamefont {{Shen}}\ \emph {et~al.}(2018)\citenamefont {{Shen}},
  \citenamefont {{Lin}}, \citenamefont {{Wang}}, \citenamefont {{Park}},
  \citenamefont {{Leong}}, \citenamefont {{Lu}}, \citenamefont {{Palacios}},\
  and\ \citenamefont {{Kong}}}]{fabrication_review_cvd}%
  \BibitemOpen
  \bibfield  {author} {\bibinfo {author} {\bibfnamefont {P.}~\bibnamefont
  {{Shen}}}, \bibinfo {author} {\bibfnamefont {Y.}~\bibnamefont {{Lin}}},
  \bibinfo {author} {\bibfnamefont {H.}~\bibnamefont {{Wang}}}, \bibinfo
  {author} {\bibfnamefont {J.}~\bibnamefont {{Park}}}, \bibinfo {author}
  {\bibfnamefont {W.~S.}\ \bibnamefont {{Leong}}}, \bibinfo {author}
  {\bibfnamefont {A.}~\bibnamefont {{Lu}}}, \bibinfo {author} {\bibfnamefont
  {T.}~\bibnamefont {{Palacios}}}, \ and\ \bibinfo {author} {\bibfnamefont
  {J.}~\bibnamefont {{Kong}}},\ }\href {\doibase 10.1109/TED.2018.2866390}
  {\bibfield  {journal} {\bibinfo  {journal} {IEEE Transactions on Electron
  Devices}\ }\textbf {\bibinfo {volume} {65}},\ \bibinfo {pages} {4040}
  (\bibinfo {year} {2018})}\BibitemShut {NoStop}%
\bibitem [{\citenamefont {Liu}\ \emph {et~al.}(2019)\citenamefont {Liu},
  \citenamefont {Zhang}, \citenamefont {He}, \citenamefont {Wang},\ and\
  \citenamefont {Liu}}]{heterostructures_review}%
  \BibitemOpen
  \bibfield  {author} {\bibinfo {author} {\bibfnamefont {Y.}~\bibnamefont
  {Liu}}, \bibinfo {author} {\bibfnamefont {S.}~\bibnamefont {Zhang}}, \bibinfo
  {author} {\bibfnamefont {J.}~\bibnamefont {He}}, \bibinfo {author}
  {\bibfnamefont {Z.~M.}\ \bibnamefont {Wang}}, \ and\ \bibinfo {author}
  {\bibfnamefont {Z.}~\bibnamefont {Liu}},\ }\href {\doibase
  10.1007/s40820-019-0245-5} {\bibfield  {journal} {\bibinfo  {journal}
  {Nano-Micro Letters}\ }\textbf {\bibinfo {volume} {11}},\ \bibinfo {pages}
  {13} (\bibinfo {year} {2019})}\BibitemShut {NoStop}%
\bibitem [{\citenamefont {Bedell}\ \emph {et~al.}(2014)\citenamefont {Bedell},
  \citenamefont {Khakifirooz},\ and\ \citenamefont
  {Sadana}}]{strainengineering_cmos}%
  \BibitemOpen
  \bibfield  {author} {\bibinfo {author} {\bibfnamefont {S.}~\bibnamefont
  {Bedell}}, \bibinfo {author} {\bibfnamefont {A.}~\bibnamefont {Khakifirooz}},
  \ and\ \bibinfo {author} {\bibfnamefont {D.}~\bibnamefont {Sadana}},\ }\href
  {\doibase 10.1557/mrs.2014.5} {\bibfield  {journal} {\bibinfo  {journal} {MRS
  Bulletin}\ }\textbf {\bibinfo {volume} {39}},\ \bibinfo {pages} {131–137}
  (\bibinfo {year} {2014})}\BibitemShut {NoStop}%
\bibitem [{\citenamefont {Dai}\ \emph {et~al.}()\citenamefont {Dai},
  \citenamefont {Liu},\ and\ \citenamefont {Zhang}}]{strainengineering_dai}%
  \BibitemOpen
  \bibfield  {author} {\bibinfo {author} {\bibfnamefont {Z.}~\bibnamefont
  {Dai}}, \bibinfo {author} {\bibfnamefont {L.}~\bibnamefont {Liu}}, \ and\
  \bibinfo {author} {\bibfnamefont {Z.}~\bibnamefont {Zhang}},\ }\href
  {\doibase 10.1002/adma.201805417} {\bibfield  {journal} {\bibinfo  {journal}
  {Advanced Materials}\ }\textbf {\bibinfo {volume} {0}},\ \bibinfo {pages}
  {1805417}},\ \Eprint
  {http://arxiv.org/abs/https://onlinelibrary.wiley.com/doi/pdf/10.1002/adma.201805417}
  {https://onlinelibrary.wiley.com/doi/pdf/10.1002/adma.201805417} \BibitemShut
  {NoStop}%
\bibitem [{\citenamefont {Akinwande}\ \emph {et~al.}(2017)\citenamefont
  {Akinwande} \emph {et~al.}}]{strainengineering_akinwande}%
  \BibitemOpen
  \bibfield  {author} {\bibinfo {author} {\bibfnamefont {D.}~\bibnamefont
  {Akinwande}} \emph {et~al.},\ }\href {\doibase
  https://doi.org/10.1016/j.eml.2017.01.008} {\bibfield  {journal} {\bibinfo
  {journal} {Extreme Mechanics Letters}\ }\textbf {\bibinfo {volume} {13}},\
  \bibinfo {pages} {42 } (\bibinfo {year} {2017})}\BibitemShut {NoStop}%
\bibitem [{\citenamefont {Pérez~Garza}\ \emph {et~al.}(2014)\citenamefont
  {Pérez~Garza}, \citenamefont {Kievit}, \citenamefont {Schneider},\ and\
  \citenamefont {Staufer}}]{graphene_garza}%
  \BibitemOpen
  \bibfield  {author} {\bibinfo {author} {\bibfnamefont {H.~H.}\ \bibnamefont
  {Pérez~Garza}}, \bibinfo {author} {\bibfnamefont {E.~W.}\ \bibnamefont
  {Kievit}}, \bibinfo {author} {\bibfnamefont {G.~F.}\ \bibnamefont
  {Schneider}}, \ and\ \bibinfo {author} {\bibfnamefont {U.}~\bibnamefont
  {Staufer}},\ }\href {\doibase 10.1021/nl5016848} {\bibfield  {journal}
  {\bibinfo  {journal} {Nano Letters}\ }\textbf {\bibinfo {volume} {14}},\
  \bibinfo {pages} {4107} (\bibinfo {year} {2014})},\ \bibinfo {note} {pMID:
  24872014},\ \Eprint {http://arxiv.org/abs/https://doi.org/10.1021/nl5016848}
  {https://doi.org/10.1021/nl5016848} \BibitemShut {NoStop}%
\bibitem [{\citenamefont {Winkler}\ and\ \citenamefont
  {Z\"ulicke}(2010)}]{graphene_winkler}%
  \BibitemOpen
  \bibfield  {author} {\bibinfo {author} {\bibfnamefont {R.}~\bibnamefont
  {Winkler}}\ and\ \bibinfo {author} {\bibfnamefont {U.}~\bibnamefont
  {Z\"ulicke}},\ }\href {\doibase 10.1103/PhysRevB.82.245313} {\bibfield
  {journal} {\bibinfo  {journal} {Phys. Rev. B}\ }\textbf {\bibinfo {volume}
  {82}},\ \bibinfo {pages} {245313} (\bibinfo {year} {2010})}\BibitemShut
  {NoStop}%
\bibitem [{\citenamefont {Cocco}\ \emph {et~al.}(2010)\citenamefont {Cocco},
  \citenamefont {Cadelano},\ and\ \citenamefont {Colombo}}]{graphene_cocco}%
  \BibitemOpen
  \bibfield  {author} {\bibinfo {author} {\bibfnamefont {G.}~\bibnamefont
  {Cocco}}, \bibinfo {author} {\bibfnamefont {E.}~\bibnamefont {Cadelano}}, \
  and\ \bibinfo {author} {\bibfnamefont {L.}~\bibnamefont {Colombo}},\ }\href
  {\doibase 10.1103/PhysRevB.81.241412} {\bibfield  {journal} {\bibinfo
  {journal} {Phys. Rev. B}\ }\textbf {\bibinfo {volume} {81}},\ \bibinfo
  {pages} {241412} (\bibinfo {year} {2010})}\BibitemShut {NoStop}%
\bibitem [{\citenamefont {Yan}\ \emph {et~al.}(2012)\citenamefont {Yan},
  \citenamefont {Sun}, \citenamefont {He}, \citenamefont {Nie},\ and\
  \citenamefont {Chan}}]{graphene_yan}%
  \BibitemOpen
  \bibfield  {author} {\bibinfo {author} {\bibfnamefont {H.}~\bibnamefont
  {Yan}}, \bibinfo {author} {\bibfnamefont {Y.}~\bibnamefont {Sun}}, \bibinfo
  {author} {\bibfnamefont {L.}~\bibnamefont {He}}, \bibinfo {author}
  {\bibfnamefont {J.-C.}\ \bibnamefont {Nie}}, \ and\ \bibinfo {author}
  {\bibfnamefont {M.~H.~W.}\ \bibnamefont {Chan}},\ }\href {\doibase
  10.1103/PhysRevB.85.035422} {\bibfield  {journal} {\bibinfo  {journal} {Phys.
  Rev. B}\ }\textbf {\bibinfo {volume} {85}},\ \bibinfo {pages} {035422}
  (\bibinfo {year} {2012})}\BibitemShut {NoStop}%
\bibitem [{\citenamefont {Li}\ \emph {et~al.}(2015)\citenamefont {Li},
  \citenamefont {Bai}, \citenamefont {Yin}, \citenamefont {Qiao}, \citenamefont
  {Wang},\ and\ \citenamefont {He}}]{graphene_li}%
  \BibitemOpen
  \bibfield  {author} {\bibinfo {author} {\bibfnamefont {S.-Y.}\ \bibnamefont
  {Li}}, \bibinfo {author} {\bibfnamefont {K.-K.}\ \bibnamefont {Bai}},
  \bibinfo {author} {\bibfnamefont {L.-J.}\ \bibnamefont {Yin}}, \bibinfo
  {author} {\bibfnamefont {J.-B.}\ \bibnamefont {Qiao}}, \bibinfo {author}
  {\bibfnamefont {W.-X.}\ \bibnamefont {Wang}}, \ and\ \bibinfo {author}
  {\bibfnamefont {L.}~\bibnamefont {He}},\ }\href {\doibase
  10.1103/PhysRevB.92.245302} {\bibfield  {journal} {\bibinfo  {journal} {Phys.
  Rev. B}\ }\textbf {\bibinfo {volume} {92}},\ \bibinfo {pages} {245302}
  (\bibinfo {year} {2015})}\BibitemShut {NoStop}%
\bibitem [{\citenamefont {Lee}\ \emph {et~al.}(2010)\citenamefont {Lee},
  \citenamefont {Bae}, \citenamefont {Jang}, \citenamefont {Jang},
  \citenamefont {Zhu}, \citenamefont {Sim}, \citenamefont {Song}, \citenamefont
  {Hong},\ and\ \citenamefont {Ahn}}]{graphene_lee}%
  \BibitemOpen
  \bibfield  {author} {\bibinfo {author} {\bibfnamefont {Y.}~\bibnamefont
  {Lee}}, \bibinfo {author} {\bibfnamefont {S.}~\bibnamefont {Bae}}, \bibinfo
  {author} {\bibfnamefont {H.}~\bibnamefont {Jang}}, \bibinfo {author}
  {\bibfnamefont {S.}~\bibnamefont {Jang}}, \bibinfo {author} {\bibfnamefont
  {S.-E.}\ \bibnamefont {Zhu}}, \bibinfo {author} {\bibfnamefont {S.~H.}\
  \bibnamefont {Sim}}, \bibinfo {author} {\bibfnamefont {Y.~I.}\ \bibnamefont
  {Song}}, \bibinfo {author} {\bibfnamefont {B.~H.}\ \bibnamefont {Hong}}, \
  and\ \bibinfo {author} {\bibfnamefont {J.-H.}\ \bibnamefont {Ahn}},\ }\href
  {\doibase 10.1021/nl903272n} {\bibfield  {journal} {\bibinfo  {journal} {Nano
  Letters}\ }\textbf {\bibinfo {volume} {10}},\ \bibinfo {pages} {490}
  (\bibinfo {year} {2010})},\ \bibinfo {note} {pMID: 20044841},\ \Eprint
  {http://arxiv.org/abs/https://doi.org/10.1021/nl903272n}
  {https://doi.org/10.1021/nl903272n} \BibitemShut {NoStop}%
\bibitem [{\citenamefont {Wang}\ \emph {et~al.}(2015)\citenamefont {Wang},
  \citenamefont {Yang}, \citenamefont {Lao}, \citenamefont {Zhang},
  \citenamefont {Zhang}, \citenamefont {Zhu}, \citenamefont {Li}, \citenamefont
  {Zang}, \citenamefont {Wang}, \citenamefont {Yu}, \citenamefont {Jin},
  \citenamefont {Wang},\ and\ \citenamefont {Zhu}}]{graphene_wang}%
  \BibitemOpen
  \bibfield  {author} {\bibinfo {author} {\bibfnamefont {Y.}~\bibnamefont
  {Wang}}, \bibinfo {author} {\bibfnamefont {T.}~\bibnamefont {Yang}}, \bibinfo
  {author} {\bibfnamefont {J.}~\bibnamefont {Lao}}, \bibinfo {author}
  {\bibfnamefont {R.}~\bibnamefont {Zhang}}, \bibinfo {author} {\bibfnamefont
  {Y.}~\bibnamefont {Zhang}}, \bibinfo {author} {\bibfnamefont
  {M.}~\bibnamefont {Zhu}}, \bibinfo {author} {\bibfnamefont {X.}~\bibnamefont
  {Li}}, \bibinfo {author} {\bibfnamefont {X.}~\bibnamefont {Zang}}, \bibinfo
  {author} {\bibfnamefont {K.}~\bibnamefont {Wang}}, \bibinfo {author}
  {\bibfnamefont {W.}~\bibnamefont {Yu}}, \bibinfo {author} {\bibfnamefont
  {H.}~\bibnamefont {Jin}}, \bibinfo {author} {\bibfnamefont {L.}~\bibnamefont
  {Wang}}, \ and\ \bibinfo {author} {\bibfnamefont {H.}~\bibnamefont {Zhu}},\
  }\href {\doibase 10.1007/s12274-014-0652-3} {\bibfield  {journal} {\bibinfo
  {journal} {Nano Research}\ }\textbf {\bibinfo {volume} {8}},\ \bibinfo
  {pages} {1627} (\bibinfo {year} {2015})}\BibitemShut {NoStop}%
\bibitem [{\citenamefont {Yang}\ \emph {et~al.}(2018)\citenamefont {Yang},
  \citenamefont {Wang}, \citenamefont {Pang}, \citenamefont {Li}, \citenamefont
  {Wang}, \citenamefont {Zhang}, \citenamefont {Wang}, \citenamefont {Liu},
  \citenamefont {Yang}, \citenamefont {Jian}, \citenamefont {Jian},
  \citenamefont {Zhang}, \citenamefont {Yang},\ and\ \citenamefont
  {Ren}}]{graphene_yang}%
  \BibitemOpen
  \bibfield  {author} {\bibinfo {author} {\bibfnamefont {Z.}~\bibnamefont
  {Yang}}, \bibinfo {author} {\bibfnamefont {D.-Y.}\ \bibnamefont {Wang}},
  \bibinfo {author} {\bibfnamefont {Y.}~\bibnamefont {Pang}}, \bibinfo {author}
  {\bibfnamefont {Y.-X.}\ \bibnamefont {Li}}, \bibinfo {author} {\bibfnamefont
  {Q.}~\bibnamefont {Wang}}, \bibinfo {author} {\bibfnamefont {T.-Y.}\
  \bibnamefont {Zhang}}, \bibinfo {author} {\bibfnamefont {J.-B.}\ \bibnamefont
  {Wang}}, \bibinfo {author} {\bibfnamefont {X.}~\bibnamefont {Liu}}, \bibinfo
  {author} {\bibfnamefont {Y.-Y.}\ \bibnamefont {Yang}}, \bibinfo {author}
  {\bibfnamefont {J.-M.}\ \bibnamefont {Jian}}, \bibinfo {author}
  {\bibfnamefont {M.-Q.}\ \bibnamefont {Jian}}, \bibinfo {author}
  {\bibfnamefont {Y.-Y.}\ \bibnamefont {Zhang}}, \bibinfo {author}
  {\bibfnamefont {Y.}~\bibnamefont {Yang}}, \ and\ \bibinfo {author}
  {\bibfnamefont {T.-L.}\ \bibnamefont {Ren}},\ }\href {\doibase
  10.1021/acsami.7b16284} {\bibfield  {journal} {\bibinfo  {journal} {ACS
  Applied Materials \& Interfaces}\ }\textbf {\bibinfo {volume} {10}},\
  \bibinfo {pages} {3948} (\bibinfo {year} {2018})},\ \bibinfo {note} {pMID:
  29281246},\ \Eprint
  {http://arxiv.org/abs/https://doi.org/10.1021/acsami.7b16284}
  {https://doi.org/10.1021/acsami.7b16284} \BibitemShut {NoStop}%
\bibitem [{\citenamefont {Voon}\ \emph {et~al.}(2015)\citenamefont {Voon},
  \citenamefont {Lopez-Bezanilla}, \citenamefont {Wang}, \citenamefont
  {Zhang},\ and\ \citenamefont {Willatzen}}]{WillatzenNJP}%
  \BibitemOpen
  \bibfield  {author} {\bibinfo {author} {\bibfnamefont {L.~C. L.~Y.}\
  \bibnamefont {Voon}}, \bibinfo {author} {\bibfnamefont {A.}~\bibnamefont
  {Lopez-Bezanilla}}, \bibinfo {author} {\bibfnamefont {J.}~\bibnamefont
  {Wang}}, \bibinfo {author} {\bibfnamefont {Y.}~\bibnamefont {Zhang}}, \ and\
  \bibinfo {author} {\bibfnamefont {M.}~\bibnamefont {Willatzen}},\ }\href
  {\doibase 10.1088/1367-2630/17/2/025004} {\bibfield  {journal} {\bibinfo
  {journal} {New J. Phys.}\ }\textbf {\bibinfo {volume} {17}},\ \bibinfo
  {pages} {025004} (\bibinfo {year} {2015})}\BibitemShut {NoStop}%
\bibitem [{\citenamefont {Xiao}\ \emph {et~al.}(2012)\citenamefont {Xiao},
  \citenamefont {Liu}, \citenamefont {Feng}, \citenamefont {Xu},\ and\
  \citenamefont {Yao}}]{MoS2_xiao}%
  \BibitemOpen
  \bibfield  {author} {\bibinfo {author} {\bibfnamefont {D.}~\bibnamefont
  {Xiao}}, \bibinfo {author} {\bibfnamefont {G.-B.}\ \bibnamefont {Liu}},
  \bibinfo {author} {\bibfnamefont {W.}~\bibnamefont {Feng}}, \bibinfo {author}
  {\bibfnamefont {X.}~\bibnamefont {Xu}}, \ and\ \bibinfo {author}
  {\bibfnamefont {W.}~\bibnamefont {Yao}},\ }\href {\doibase
  10.1103/PhysRevLett.108.196802} {\bibfield  {journal} {\bibinfo  {journal}
  {Phys. Rev. Lett.}\ }\textbf {\bibinfo {volume} {108}},\ \bibinfo {pages}
  {196802} (\bibinfo {year} {2012})}\BibitemShut {NoStop}%
\bibitem [{\citenamefont {Beiranvand}\ \emph {et~al.}(2018)\citenamefont
  {Beiranvand}, \citenamefont {Dezfuli},\ and\ \citenamefont
  {Sabaeian}}]{MoS2_bereinvand}%
  \BibitemOpen
  \bibfield  {author} {\bibinfo {author} {\bibfnamefont {K.}~\bibnamefont
  {Beiranvand}}, \bibinfo {author} {\bibfnamefont {A.~G.}\ \bibnamefont
  {Dezfuli}}, \ and\ \bibinfo {author} {\bibfnamefont {M.}~\bibnamefont
  {Sabaeian}},\ }\href {\doibase https://doi.org/10.1016/j.spmi.2018.06.033}
  {\bibfield  {journal} {\bibinfo  {journal} {Superlattices and
  Microstructures}\ }\textbf {\bibinfo {volume} {120}},\ \bibinfo {pages} {812
  } (\bibinfo {year} {2018})}\BibitemShut {NoStop}%
\bibitem [{\citenamefont {Korm\'anyos}\ \emph {et~al.}(2013)\citenamefont
  {Korm\'anyos}, \citenamefont {Z\'olyomi}, \citenamefont {Drummond},
  \citenamefont {Rakyta}, \citenamefont {Burkard},\ and\ \citenamefont
  {Fal'ko}}]{MoS2_kormanyos}%
  \BibitemOpen
  \bibfield  {author} {\bibinfo {author} {\bibfnamefont {A.}~\bibnamefont
  {Korm\'anyos}}, \bibinfo {author} {\bibfnamefont {V.}~\bibnamefont
  {Z\'olyomi}}, \bibinfo {author} {\bibfnamefont {N.~D.}\ \bibnamefont
  {Drummond}}, \bibinfo {author} {\bibfnamefont {P.}~\bibnamefont {Rakyta}},
  \bibinfo {author} {\bibfnamefont {G.}~\bibnamefont {Burkard}}, \ and\
  \bibinfo {author} {\bibfnamefont {V.~I.}\ \bibnamefont {Fal'ko}},\ }\href
  {\doibase 10.1103/PhysRevB.88.045416} {\bibfield  {journal} {\bibinfo
  {journal} {Phys. Rev. B}\ }\textbf {\bibinfo {volume} {88}},\ \bibinfo
  {pages} {045416} (\bibinfo {year} {2013})}\BibitemShut {NoStop}%
\bibitem [{\citenamefont {He}\ \emph {et~al.}(2016)\citenamefont {He},
  \citenamefont {Li}, \citenamefont {Zhu}, \citenamefont {Dai}, \citenamefont
  {Yang}, \citenamefont {Yang}, \citenamefont {Zhang}, \citenamefont {Li},
  \citenamefont {Schwingenschlogl},\ and\ \citenamefont {Zhang}}]{MoS2_he}%
  \BibitemOpen
  \bibfield  {author} {\bibinfo {author} {\bibfnamefont {X.}~\bibnamefont
  {He}}, \bibinfo {author} {\bibfnamefont {H.}~\bibnamefont {Li}}, \bibinfo
  {author} {\bibfnamefont {Z.}~\bibnamefont {Zhu}}, \bibinfo {author}
  {\bibfnamefont {Z.}~\bibnamefont {Dai}}, \bibinfo {author} {\bibfnamefont
  {Y.}~\bibnamefont {Yang}}, \bibinfo {author} {\bibfnamefont {P.}~\bibnamefont
  {Yang}}, \bibinfo {author} {\bibfnamefont {Q.}~\bibnamefont {Zhang}},
  \bibinfo {author} {\bibfnamefont {P.}~\bibnamefont {Li}}, \bibinfo {author}
  {\bibfnamefont {U.}~\bibnamefont {Schwingenschlogl}}, \ and\ \bibinfo
  {author} {\bibfnamefont {X.}~\bibnamefont {Zhang}},\ }\href {\doibase
  10.1063/1.4966218} {\bibfield  {journal} {\bibinfo  {journal} {Applied
  Physics Letters}\ }\textbf {\bibinfo {volume} {109}},\ \bibinfo {pages}
  {173105} (\bibinfo {year} {2016})},\ \Eprint
  {http://arxiv.org/abs/https://doi.org/10.1063/1.4966218}
  {https://doi.org/10.1063/1.4966218} \BibitemShut {NoStop}%
\bibitem [{\citenamefont {Schmidt}\ \emph {et~al.}(2016)\citenamefont
  {Schmidt}, \citenamefont {Niehues}, \citenamefont {Schneider}, \citenamefont
  {Drüppel}, \citenamefont {Deilmann}, \citenamefont {Rohlfing}, \citenamefont
  {de~Vasconcellos}, \citenamefont {Castellanos-Gomez},\ and\ \citenamefont
  {Bratschitsch}}]{WSe2_schmidt}%
  \BibitemOpen
  \bibfield  {author} {\bibinfo {author} {\bibfnamefont {R.}~\bibnamefont
  {Schmidt}}, \bibinfo {author} {\bibfnamefont {I.}~\bibnamefont {Niehues}},
  \bibinfo {author} {\bibfnamefont {R.}~\bibnamefont {Schneider}}, \bibinfo
  {author} {\bibfnamefont {M.}~\bibnamefont {Drüppel}}, \bibinfo {author}
  {\bibfnamefont {T.}~\bibnamefont {Deilmann}}, \bibinfo {author}
  {\bibfnamefont {M.}~\bibnamefont {Rohlfing}}, \bibinfo {author}
  {\bibfnamefont {S.~M.}\ \bibnamefont {de~Vasconcellos}}, \bibinfo {author}
  {\bibfnamefont {A.}~\bibnamefont {Castellanos-Gomez}}, \ and\ \bibinfo
  {author} {\bibfnamefont {R.}~\bibnamefont {Bratschitsch}},\ }\href {\doibase
  10.1088/2053-1583/3/2/021011} {\bibfield  {journal} {\bibinfo  {journal} {2D
  Materials}\ }\textbf {\bibinfo {volume} {3}},\ \bibinfo {pages} {021011}
  (\bibinfo {year} {2016})}\BibitemShut {NoStop}%
\bibitem [{\citenamefont {Conley}\ \emph {et~al.}(2013)\citenamefont {Conley},
  \citenamefont {Wang}, \citenamefont {Ziegler}, \citenamefont {Haglund},
  \citenamefont {Pantelides},\ and\ \citenamefont {Bolotin}}]{MoS2_conley}%
  \BibitemOpen
  \bibfield  {author} {\bibinfo {author} {\bibfnamefont {H.~J.}\ \bibnamefont
  {Conley}}, \bibinfo {author} {\bibfnamefont {B.}~\bibnamefont {Wang}},
  \bibinfo {author} {\bibfnamefont {J.~I.}\ \bibnamefont {Ziegler}}, \bibinfo
  {author} {\bibfnamefont {R.~F.}\ \bibnamefont {Haglund}}, \bibinfo {author}
  {\bibfnamefont {S.~T.}\ \bibnamefont {Pantelides}}, \ and\ \bibinfo {author}
  {\bibfnamefont {K.~I.}\ \bibnamefont {Bolotin}},\ }\href {\doibase
  10.1021/nl4014748} {\bibfield  {journal} {\bibinfo  {journal} {Nano Letters}\
  }\textbf {\bibinfo {volume} {13}},\ \bibinfo {pages} {3626} (\bibinfo {year}
  {2013})},\ \bibinfo {note} {pMID: 23819588},\ \Eprint
  {http://arxiv.org/abs/https://doi.org/10.1021/nl4014748}
  {https://doi.org/10.1021/nl4014748} \BibitemShut {NoStop}%
\bibitem [{\citenamefont {Wu}\ \emph {et~al.}(2018)\citenamefont {Wu},
  \citenamefont {Wang}, \citenamefont {Ercius}, \citenamefont {Wright},
  \citenamefont {Leppert-Simenauer}, \citenamefont {Burke}, \citenamefont
  {Dubey}, \citenamefont {Dogare},\ and\ \citenamefont {Pettes}}]{WSe2_wu}%
  \BibitemOpen
  \bibfield  {author} {\bibinfo {author} {\bibfnamefont {W.}~\bibnamefont
  {Wu}}, \bibinfo {author} {\bibfnamefont {J.}~\bibnamefont {Wang}}, \bibinfo
  {author} {\bibfnamefont {P.}~\bibnamefont {Ercius}}, \bibinfo {author}
  {\bibfnamefont {N.~C.}\ \bibnamefont {Wright}}, \bibinfo {author}
  {\bibfnamefont {D.~M.}\ \bibnamefont {Leppert-Simenauer}}, \bibinfo {author}
  {\bibfnamefont {R.~A.}\ \bibnamefont {Burke}}, \bibinfo {author}
  {\bibfnamefont {M.}~\bibnamefont {Dubey}}, \bibinfo {author} {\bibfnamefont
  {A.~M.}\ \bibnamefont {Dogare}}, \ and\ \bibinfo {author} {\bibfnamefont
  {M.~T.}\ \bibnamefont {Pettes}},\ }\href {\doibase
  10.1021/acs.nanolett.7b05229} {\bibfield  {journal} {\bibinfo  {journal}
  {Nano Letters}\ }\textbf {\bibinfo {volume} {18}},\ \bibinfo {pages} {2351}
  (\bibinfo {year} {2018})},\ \bibinfo {note} {pMID: 29558623},\ \Eprint
  {http://arxiv.org/abs/https://doi.org/10.1021/acs.nanolett.7b05229}
  {https://doi.org/10.1021/acs.nanolett.7b05229} \BibitemShut {NoStop}%
\bibitem [{\citenamefont {Castellanos-Gomez}\ \emph {et~al.}(2013)\citenamefont
  {Castellanos-Gomez}, \citenamefont {Roldán}, \citenamefont {Cappelluti},
  \citenamefont {Buscema}, \citenamefont {Guinea}, \citenamefont {van~der
  Zant},\ and\ \citenamefont {Steele}}]{MoS2_gomez}%
  \BibitemOpen
  \bibfield  {author} {\bibinfo {author} {\bibfnamefont {A.}~\bibnamefont
  {Castellanos-Gomez}}, \bibinfo {author} {\bibfnamefont {R.}~\bibnamefont
  {Roldán}}, \bibinfo {author} {\bibfnamefont {E.}~\bibnamefont {Cappelluti}},
  \bibinfo {author} {\bibfnamefont {M.}~\bibnamefont {Buscema}}, \bibinfo
  {author} {\bibfnamefont {F.}~\bibnamefont {Guinea}}, \bibinfo {author}
  {\bibfnamefont {H.~S.~J.}\ \bibnamefont {van~der Zant}}, \ and\ \bibinfo
  {author} {\bibfnamefont {G.~A.}\ \bibnamefont {Steele}},\ }\href {\doibase
  10.1021/nl402875m} {\bibfield  {journal} {\bibinfo  {journal} {Nano Letters}\
  }\textbf {\bibinfo {volume} {13}},\ \bibinfo {pages} {5361} (\bibinfo {year}
  {2013})},\ \bibinfo {note} {pMID: 24083520},\ \Eprint
  {http://arxiv.org/abs/https://doi.org/10.1021/nl402875m}
  {https://doi.org/10.1021/nl402875m} \BibitemShut {NoStop}%
\bibitem [{\citenamefont {{Gant}}\ \emph {et~al.}(2019)\citenamefont {{Gant}},
  \citenamefont {{Huang}}, \citenamefont {{P{\'e}rez de Lara}}, \citenamefont
  {{Guo}}, \citenamefont {{Frisenda}},\ and\ \citenamefont
  {{Castellanos-Gomez}}}]{MoS2_gant}%
  \BibitemOpen
  \bibfield  {author} {\bibinfo {author} {\bibfnamefont {P.}~\bibnamefont
  {{Gant}}}, \bibinfo {author} {\bibfnamefont {P.}~\bibnamefont {{Huang}}},
  \bibinfo {author} {\bibfnamefont {D.}~\bibnamefont {{P{\'e}rez de Lara}}},
  \bibinfo {author} {\bibfnamefont {D.}~\bibnamefont {{Guo}}}, \bibinfo
  {author} {\bibfnamefont {R.}~\bibnamefont {{Frisenda}}}, \ and\ \bibinfo
  {author} {\bibfnamefont {A.}~\bibnamefont {{Castellanos-Gomez}}},\
  }\href@noop {} {\bibfield  {journal} {\bibinfo  {journal} {arXiv e-prints}\
  ,\ \bibinfo {eid} {arXiv:1902.02802}} (\bibinfo {year} {2019})},\ \Eprint
  {http://arxiv.org/abs/1902.02802} {arXiv:1902.02802 [physics.app-ph]}
  \BibitemShut {NoStop}%
\bibitem [{\citenamefont {Zubair}\ \emph {et~al.}(2017)\citenamefont {Zubair},
  \citenamefont {Tahir}, \citenamefont {Vasilopoulos},\ and\ \citenamefont
  {Sabeeh}}]{MoS2_zubair}%
  \BibitemOpen
  \bibfield  {author} {\bibinfo {author} {\bibfnamefont {M.}~\bibnamefont
  {Zubair}}, \bibinfo {author} {\bibfnamefont {M.}~\bibnamefont {Tahir}},
  \bibinfo {author} {\bibfnamefont {P.}~\bibnamefont {Vasilopoulos}}, \ and\
  \bibinfo {author} {\bibfnamefont {K.}~\bibnamefont {Sabeeh}},\ }\href
  {\doibase 10.1103/PhysRevB.96.045405} {\bibfield  {journal} {\bibinfo
  {journal} {Phys. Rev. B}\ }\textbf {\bibinfo {volume} {96}},\ \bibinfo
  {pages} {045405} (\bibinfo {year} {2017})}\BibitemShut {NoStop}%
\bibitem [{\citenamefont {Qi}\ \emph {et~al.}(2013)\citenamefont {Qi},
  \citenamefont {Li}, \citenamefont {Qian},\ and\ \citenamefont
  {Feng}}]{MoS2_qi}%
  \BibitemOpen
  \bibfield  {author} {\bibinfo {author} {\bibfnamefont {J.}~\bibnamefont
  {Qi}}, \bibinfo {author} {\bibfnamefont {X.}~\bibnamefont {Li}}, \bibinfo
  {author} {\bibfnamefont {X.}~\bibnamefont {Qian}}, \ and\ \bibinfo {author}
  {\bibfnamefont {J.}~\bibnamefont {Feng}},\ }\href {\doibase
  10.1063/1.4803803} {\bibfield  {journal} {\bibinfo  {journal} {Applied
  Physics Letters}\ }\textbf {\bibinfo {volume} {102}},\ \bibinfo {pages}
  {173112} (\bibinfo {year} {2013})},\ \Eprint
  {http://arxiv.org/abs/https://doi.org/10.1063/1.4803803}
  {https://doi.org/10.1063/1.4803803} \BibitemShut {NoStop}%
\bibitem [{\citenamefont {Manoharan}\ and\ \citenamefont
  {Subramanian}(2018)}]{BAs}%
  \BibitemOpen
  \bibfield  {author} {\bibinfo {author} {\bibfnamefont {K.}~\bibnamefont
  {Manoharan}}\ and\ \bibinfo {author} {\bibfnamefont {V.}~\bibnamefont
  {Subramanian}},\ }\href {\doibase 10.1021/acsomega.8b00946} {\bibfield
  {journal} {\bibinfo  {journal} {ACS Omega}\ }\textbf {\bibinfo {volume}
  {3}},\ \bibinfo {pages} {9533} (\bibinfo {year} {2018})},\ \Eprint
  {http://arxiv.org/abs/https://doi.org/10.1021/acsomega.8b00946}
  {https://doi.org/10.1021/acsomega.8b00946} \BibitemShut {NoStop}%
\bibitem [{\citenamefont {\ifmmode~\mbox{\c{S}}\else \c{S}\fi{}ahin}\ \emph
  {et~al.}(2009)\citenamefont {\ifmmode~\mbox{\c{S}}\else \c{S}\fi{}ahin},
  \citenamefont {Cahangirov}, \citenamefont {Topsakal}, \citenamefont
  {Bekaroglu}, \citenamefont {Akturk}, \citenamefont {Senger},\ and\
  \citenamefont {Ciraci}}]{BAs_sahin}%
  \BibitemOpen
  \bibfield  {author} {\bibinfo {author} {\bibfnamefont {H.}~\bibnamefont
  {\ifmmode~\mbox{\c{S}}\else \c{S}\fi{}ahin}}, \bibinfo {author}
  {\bibfnamefont {S.}~\bibnamefont {Cahangirov}}, \bibinfo {author}
  {\bibfnamefont {M.}~\bibnamefont {Topsakal}}, \bibinfo {author}
  {\bibfnamefont {E.}~\bibnamefont {Bekaroglu}}, \bibinfo {author}
  {\bibfnamefont {E.}~\bibnamefont {Akturk}}, \bibinfo {author} {\bibfnamefont
  {R.~T.}\ \bibnamefont {Senger}}, \ and\ \bibinfo {author} {\bibfnamefont
  {S.}~\bibnamefont {Ciraci}},\ }\href {\doibase 10.1103/PhysRevB.80.155453}
  {\bibfield  {journal} {\bibinfo  {journal} {Phys. Rev. B}\ }\textbf {\bibinfo
  {volume} {80}},\ \bibinfo {pages} {155453} (\bibinfo {year}
  {2009})}\BibitemShut {NoStop}%
\bibitem [{\citenamefont {Tian}\ \emph {et~al.}(2018)\citenamefont {Tian} \emph
  {et~al.}}]{BAs_tian}%
  \BibitemOpen
  \bibfield  {author} {\bibinfo {author} {\bibfnamefont {F.}~\bibnamefont
  {Tian}} \emph {et~al.},\ }\href {\doibase 10.1126/science.aat7932} {\bibfield
   {journal} {\bibinfo  {journal} {Science}\ }\textbf {\bibinfo {volume}
  {361}},\ \bibinfo {pages} {582} (\bibinfo {year} {2018})},\ \Eprint
  {http://arxiv.org/abs/https://science.sciencemag.org/content/361/6402/582.full.pdf}
  {https://science.sciencemag.org/content/361/6402/582.full.pdf} \BibitemShut
  {NoStop}%
\bibitem [{\citenamefont {Kang}\ \emph {et~al.}(2018)\citenamefont {Kang},
  \citenamefont {Li}, \citenamefont {Wu}, \citenamefont {Nguyen},\ and\
  \citenamefont {Hu}}]{BAs_kang}%
  \BibitemOpen
  \bibfield  {author} {\bibinfo {author} {\bibfnamefont {J.~S.}\ \bibnamefont
  {Kang}}, \bibinfo {author} {\bibfnamefont {M.}~\bibnamefont {Li}}, \bibinfo
  {author} {\bibfnamefont {H.}~\bibnamefont {Wu}}, \bibinfo {author}
  {\bibfnamefont {H.}~\bibnamefont {Nguyen}}, \ and\ \bibinfo {author}
  {\bibfnamefont {Y.}~\bibnamefont {Hu}},\ }\href {\doibase
  10.1126/science.aat5522} {\bibfield  {journal} {\bibinfo  {journal}
  {Science}\ }\textbf {\bibinfo {volume} {361}},\ \bibinfo {pages} {575}
  (\bibinfo {year} {2018})},\ \Eprint
  {http://arxiv.org/abs/https://science.sciencemag.org/content/361/6402/575.full.pdf}
  {https://science.sciencemag.org/content/361/6402/575.full.pdf} \BibitemShut
  {NoStop}%
\bibitem [{\citenamefont {{Shi}}\ and\ \citenamefont {{Luo}}(2018)}]{BAs_shi}%
  \BibitemOpen
  \bibfield  {author} {\bibinfo {author} {\bibfnamefont {C.}~\bibnamefont
  {{Shi}}}\ and\ \bibinfo {author} {\bibfnamefont {X.}~\bibnamefont {{Luo}}},\
  }\href@noop {} {\bibfield  {journal} {\bibinfo  {journal} {arXiv e-prints}\
  ,\ \bibinfo {eid} {arXiv:1811.05597}} (\bibinfo {year} {2018})},\ \Eprint
  {http://arxiv.org/abs/1811.05597} {arXiv:1811.05597 [cond-mat.mtrl-sci]}
  \BibitemShut {NoStop}%
\bibitem [{\citenamefont {Ren}\ \emph {et~al.}(2019)\citenamefont {Ren},
  \citenamefont {Kong},\ and\ \citenamefont {Ni}}]{BAs_ren}%
  \BibitemOpen
  \bibfield  {author} {\bibinfo {author} {\bibfnamefont {J.}~\bibnamefont
  {Ren}}, \bibinfo {author} {\bibfnamefont {W.}~\bibnamefont {Kong}}, \ and\
  \bibinfo {author} {\bibfnamefont {J.}~\bibnamefont {Ni}},\ }\href {\doibase
  10.1186/s11671-019-2972-4} {\bibfield  {journal} {\bibinfo  {journal}
  {Nanoscale Research Letters}\ }\textbf {\bibinfo {volume} {14}},\ \bibinfo
  {pages} {133} (\bibinfo {year} {2019})}\BibitemShut {NoStop}%
\bibitem [{\citenamefont {wu~Zhang}\ \emph {et~al.}(2015)\citenamefont
  {wu~Zhang}, \citenamefont {wen Zhang}, \citenamefont {xiao Ji}, \citenamefont
  {shi Li}, \citenamefont {ji~Wang}, \citenamefont {jun Hu},\ and\
  \citenamefont {shen Yan}}]{BAs_zhang}%
  \BibitemOpen
  \bibfield  {author} {\bibinfo {author} {\bibfnamefont {R.}~\bibnamefont
  {wu~Zhang}}, \bibinfo {author} {\bibfnamefont {C.}~\bibnamefont {wen Zhang}},
  \bibinfo {author} {\bibfnamefont {W.}~\bibnamefont {xiao Ji}}, \bibinfo
  {author} {\bibfnamefont {S.}~\bibnamefont {shi Li}}, \bibinfo {author}
  {\bibfnamefont {P.}~\bibnamefont {ji~Wang}}, \bibinfo {author} {\bibfnamefont
  {S.}~\bibnamefont {jun Hu}}, \ and\ \bibinfo {author} {\bibfnamefont
  {S.}~\bibnamefont {shen Yan}},\ }\href {\doibase 10.7567/apex.8.113001}
  {\bibfield  {journal} {\bibinfo  {journal} {Applied Physics Express}\
  }\textbf {\bibinfo {volume} {8}},\ \bibinfo {pages} {113001} (\bibinfo {year}
  {2015})}\BibitemShut {NoStop}%
\bibitem [{\citenamefont {Ullah}\ \emph {et~al.}(2018)\citenamefont {Ullah},
  \citenamefont {Denis},\ and\ \citenamefont {Sato}}]{BAs_ullah}%
  \BibitemOpen
  \bibfield  {author} {\bibinfo {author} {\bibfnamefont {S.}~\bibnamefont
  {Ullah}}, \bibinfo {author} {\bibfnamefont {P.~A.}\ \bibnamefont {Denis}}, \
  and\ \bibinfo {author} {\bibfnamefont {F.}~\bibnamefont {Sato}},\ }\href
  {\doibase 10.1021/acsomega.8b02605} {\bibfield  {journal} {\bibinfo
  {journal} {ACS Omega}\ }\textbf {\bibinfo {volume} {3}},\ \bibinfo {pages}
  {16416} (\bibinfo {year} {2018})},\ \Eprint
  {http://arxiv.org/abs/https://doi.org/10.1021/acsomega.8b02605}
  {https://doi.org/10.1021/acsomega.8b02605} \BibitemShut {NoStop}%
\bibitem [{\citenamefont {Lew Yan~Voon}\ and\ \citenamefont
  {Willatzen}(2009)}]{kpmethod}%
  \BibitemOpen
  \bibfield  {author} {\bibinfo {author} {\bibfnamefont {L.~C.}\ \bibnamefont
  {Lew Yan~Voon}}\ and\ \bibinfo {author} {\bibfnamefont {M.}~\bibnamefont
  {Willatzen}},\ }\href@noop {} {\emph {\bibinfo {title} {The k.p Method}}}\
  (\bibinfo  {publisher} {Springer},\ \bibinfo {year} {2009})\BibitemShut
  {NoStop}%
\bibitem [{\citenamefont {Bir}\ and\ \citenamefont {Pikus}(1974)}]{BirPikus}%
  \BibitemOpen
  \bibfield  {author} {\bibinfo {author} {\bibfnamefont {G.~L.}\ \bibnamefont
  {Bir}}\ and\ \bibinfo {author} {\bibfnamefont {G.~E.}\ \bibnamefont
  {Pikus}},\ }\href@noop {} {\emph {\bibinfo {title} {Symmetry and
  Strain-induced Effects in Semiconductors}}}\ (\bibinfo  {publisher} {John
  Wiley \& sons},\ \bibinfo {year} {1974})\BibitemShut {NoStop}%
\bibitem [{\citenamefont {Koster}\ \emph {et~al.}(1963)\citenamefont {Koster},
  \citenamefont {Dimmock}, \citenamefont {Wheeler},\ and\ \citenamefont
  {Statz}}]{koster}%
  \BibitemOpen
  \bibfield  {author} {\bibinfo {author} {\bibfnamefont {G.~F.}\ \bibnamefont
  {Koster}}, \bibinfo {author} {\bibfnamefont {J.~O.}\ \bibnamefont {Dimmock}},
  \bibinfo {author} {\bibfnamefont {R.~G.}\ \bibnamefont {Wheeler}}, \ and\
  \bibinfo {author} {\bibfnamefont {H.}~\bibnamefont {Statz}},\ }\href@noop {}
  {\emph {\bibinfo {title} {Properties of the thirty-two Point Groups}}}\
  (\bibinfo  {publisher} {M.I.T. Press},\ \bibinfo {year} {1963})\BibitemShut
  {NoStop}%
\bibitem [{\citenamefont {Dresselhaus}\ \emph {et~al.}(2008)\citenamefont
  {Dresselhaus}, \citenamefont {Dresselhaus},\ and\ \citenamefont
  {Jorio}}]{Dresselhaus}%
  \BibitemOpen
  \bibfield  {author} {\bibinfo {author} {\bibfnamefont {M.~S.}\ \bibnamefont
  {Dresselhaus}}, \bibinfo {author} {\bibfnamefont {G.}~\bibnamefont
  {Dresselhaus}}, \ and\ \bibinfo {author} {\bibfnamefont {A.}~\bibnamefont
  {Jorio}},\ }\href@noop {} {\emph {\bibinfo {title} {Group Theory -
  Application to the Physics of Condensed Matter}}}\ (\bibinfo  {publisher}
  {Springer},\ \bibinfo {year} {2008})\BibitemShut {NoStop}%
\bibitem [{\citenamefont {Mortensen}\ \emph {et~al.}(2005)\citenamefont
  {Mortensen}, \citenamefont {Hansen},\ and\ \citenamefont {Jacobsen}}]{GPAW1}%
  \BibitemOpen
  \bibfield  {author} {\bibinfo {author} {\bibfnamefont {J.~J.}\ \bibnamefont
  {Mortensen}}, \bibinfo {author} {\bibfnamefont {L.~B.}\ \bibnamefont
  {Hansen}}, \ and\ \bibinfo {author} {\bibfnamefont {K.~W.}\ \bibnamefont
  {Jacobsen}},\ }\href {\doibase 10.1103/PhysRevB.71.035109} {\bibfield
  {journal} {\bibinfo  {journal} {Phys. Rev. B}\ }\textbf {\bibinfo {volume}
  {71}},\ \bibinfo {pages} {035109} (\bibinfo {year} {2005})}\BibitemShut
  {NoStop}%
\bibitem [{\citenamefont {Enkovaara}\ \emph {et~al.}(2010)\citenamefont
  {Enkovaara} \emph {et~al.}}]{GPAW2}%
  \BibitemOpen
  \bibfield  {author} {\bibinfo {author} {\bibfnamefont {J.}~\bibnamefont
  {Enkovaara}} \emph {et~al.},\ }\href
  {http://stacks.iop.org/0953-8984/22/i=25/a=253202} {\bibfield  {journal}
  {\bibinfo  {journal} {Journal of Physics: Condensed Matter}\ }\textbf
  {\bibinfo {volume} {22}},\ \bibinfo {pages} {253202} (\bibinfo {year}
  {2010})}\BibitemShut {NoStop}%
\bibitem [{\citenamefont {Perdew}\ \emph {et~al.}(1996)\citenamefont {Perdew},
  \citenamefont {Burke},\ and\ \citenamefont {Ernzerhof}}]{PBE}%
  \BibitemOpen
  \bibfield  {author} {\bibinfo {author} {\bibfnamefont {J.~P.}\ \bibnamefont
  {Perdew}}, \bibinfo {author} {\bibfnamefont {K.}~\bibnamefont {Burke}}, \
  and\ \bibinfo {author} {\bibfnamefont {M.}~\bibnamefont {Ernzerhof}},\ }\href
  {\doibase 10.1103/PhysRevLett.77.3865} {\bibfield  {journal} {\bibinfo
  {journal} {Phys. Rev. Lett.}\ }\textbf {\bibinfo {volume} {77}},\ \bibinfo
  {pages} {3865} (\bibinfo {year} {1996})}\BibitemShut {NoStop}%
\bibitem [{\citenamefont {Larsen}\ \emph {et~al.}(2017)\citenamefont {Larsen}
  \emph {et~al.}}]{ase}%
  \BibitemOpen
  \bibfield  {author} {\bibinfo {author} {\bibfnamefont {A.~H.}\ \bibnamefont
  {Larsen}} \emph {et~al.},\ }\href
  {http://stacks.iop.org/0953-8984/29/i=27/a=273002} {\bibfield  {journal}
  {\bibinfo  {journal} {J. Phys.: Condens. Matter}\ }\textbf {\bibinfo {volume}
  {29}},\ \bibinfo {pages} {273002} (\bibinfo {year} {2017})}\BibitemShut
  {NoStop}%
\bibitem [{\citenamefont {Olsen}(2016)}]{SOC}%
  \BibitemOpen
  \bibfield  {author} {\bibinfo {author} {\bibfnamefont {T.}~\bibnamefont
  {Olsen}},\ }\href {\doibase 10.1103/PhysRevB.94.235106} {\bibfield  {journal}
  {\bibinfo  {journal} {Phys. Rev. B}\ }\textbf {\bibinfo {volume} {94}},\
  \bibinfo {pages} {235106} (\bibinfo {year} {2016})}\BibitemShut {NoStop}%
\bibitem [{\citenamefont {Zhang}\ \emph {et~al.}(2009)\citenamefont {Zhang},
  \citenamefont {Tang}, \citenamefont {Girit}, \citenamefont {Hao},
  \citenamefont {Martin}, \citenamefont {Zettl}, \citenamefont {Crommie},\ and\
  \citenamefont {Wang}}]{graphene_zhang}%
  \BibitemOpen
  \bibfield  {author} {\bibinfo {author} {\bibfnamefont {Y.}~\bibnamefont
  {Zhang}}, \bibinfo {author} {\bibfnamefont {T.-T.}\ \bibnamefont {Tang}},
  \bibinfo {author} {\bibfnamefont {C.}~\bibnamefont {Girit}}, \bibinfo
  {author} {\bibfnamefont {Z.}~\bibnamefont {Hao}}, \bibinfo {author}
  {\bibfnamefont {M.~C.}\ \bibnamefont {Martin}}, \bibinfo {author}
  {\bibfnamefont {A.}~\bibnamefont {Zettl}}, \bibinfo {author} {\bibfnamefont
  {M.~F.}\ \bibnamefont {Crommie}}, \ and\ \bibinfo {author} {\bibfnamefont
  {Y.~R. S. .~F.}\ \bibnamefont {Wang}},\ }\href@noop {} {\bibfield  {journal}
  {\bibinfo  {journal} {Nature}\ }\textbf {\bibinfo {volume} {459}},\ \bibinfo
  {pages} {820–823} (\bibinfo {year} {2009})}\BibitemShut {NoStop}%
\end{thebibliography}%
\bibliographystyle{apsrev4-1}

\end{document}